\documentclass[prl,showpacs,superscriptaddress,twocolumn]{revtex4-1}

\usepackage{graphicx}
\usepackage{amsmath}
\usepackage{amssymb}
\usepackage{hyperref}\hypersetup{colorlinks=true,linkcolor=blue,citecolor=blue,pdfstartview=FitH}
\usepackage{bm}\renewcommand{\mathbf}{\bm}\renewcommand{\vec}{\bm}

\newcommand{\figwidth}{0.95\columnwidth}
\newcommand{\largefigwidth}{0.83\columnwidth}

\begin{document}
\title{Glass-like Behavior in a System of One Dimensional Fermions after a Quantum Quench}
\author{N.~Nessi}
\author{A.~Iucci}
\affiliation{Instituto de F\'{\i}sica La Plata (IFLP) - CONICET and Departamento de F\'{\i}sica,\\
Universidad Nacional de La Plata, CC 67, 1900 La Plata, Argentina.}

\begin{abstract}
We investigate the non-equilibrium relaxation dynamics of a one dimensional system of interacting spinless fermions near the XXZ integrable point. We observe two qualitatively different regimes: close to integrability \emph{and} for low energies the relaxation proceeds in two steps (prethermalization scenario), while for large energies and/or away from integrability the dynamics develops in a single step. When the integrability breaking parameter is below a certain finite threshold and the energy of the system is sufficiently low the lifetime of the metastable states increases abruptly by several orders of magnitude, resembling the physics of glassy systems. This is reflected in a sudden jump in the relaxation timescales. We present results for finite but large systems and for large times compared to standard numerical methods. Our approach is based on the construction of equations of motion for one- and two-particle correlation functions using projection operator techniques.
\end{abstract}


\maketitle

Due to the peculiarities of quantum dynamics, it is possible to calculate expectation values of observables in the asymptotic state the system would reach after an infinite evolution without requiring the dynamics that lead, starting from a given non-equilibrium initial condition, to such state~\cite{rigol08_mechanism_thermalization}. For non-integrable systems (even infinitesimally close to an integrable point) it has been shown that one-particle observables in such asymptotic state are in very good agreement with the predictions of statistical mechanics in the thermodynamic limit~\cite{rigol14_quenches_thermo_limit}. There is also evidence that the indicators of quantum chaos related to the statistical properties of the Hamiltonian's eigenspectrum~\cite{santos10_quantum_chaos_1D} change abruptly when infinitesimally breaking integrability. However, thermalization of a closed system is a dynamical process. In particular, the dynamics toward the final thermal state in nearly integrable systems seems to be more rich and subtle than the dichotomic situation present in the static results.

In fact, it is well known that systems close to an integrable point may exhibit non-thermal stationary states, both in one dimension ($1D$) and in higher dimensions~\cite{moeckel08_quench_hubbard_high_d,eckstein09_thermalization_quench_fermi_hubbard,marino12_ising_noisy,vandenworm_2013,essler13_quench_tuneable_integrability_breaking,kollar11_gge_pretherm,stark13_kinetic_description,nessi14_gge_fermi_liquid,marcuzzi13_pretherm_ising}. Such metastable states emerge as a result of a fast lost of memory of the initial conditions due to dephasing~\cite{nessi14_gge_fermi_liquid,mitra_prethermalization_2013}. The emergence of such non-thermal stationary states has been observed experimentally in cold atomic Bose gases~\cite{gring12_prethermalization_isolated_bose_gas,langen15_quench_lieb-liniger}. Quite interestingly, dephasing alone may lead to complete thermalization of some observables, such as the kinetic energy of the system, a phenomenon dubbed prethermalization in Ref.~\cite{berges04_prethermalization_idea}. However, little is known about the subsequent evolution of the system after getting caught in such metastable states, in part due to the formidable technical challenge that poses the out-of-equilibrium dynamics of many-body quantum systems. For example, simple questions such as how much do these stationary states live or which are the relevant timescales involved in the thermalization process has no definitive answer yet.

In this work we analyze the dynamics of a system of $1D$ spinless fermions near the XXZ integrable point while varying the distance to integrability and the energy of the system. The main finding of our work is that below a certain \emph{finite} threshold distance to integrability \emph{and} for sufficiently low energies these metastable states are extremely long-lived. Their lifetime increases by several orders of magnitude while crossing such threshold. Such metastable states may completely hinder the observation of the final thermal equilibrium state in experiments or numerical simulations leading to an apparent lack of thermalization. 
 We find the situation rather similar to that of glasses, systems that exhibit the typical two-step relaxation as a consequence of getting caught in extremely long-lived metastable states, whose lifetime can be of geological scale for sufficiently low temperatures or high densities. Glass-like behavior, including ageing and slow relaxation, has also been found in open quantum systems~\cite{poletti13_glass_open_hubbard,sciolla14_ageing_open_hubbard}. Our approach, based on the projection operator formalism~\cite{grabert82_book_pot,rau95_review_pot,breuer82_book_open_systems,robertson66_pot_first,kawasaki73_non-linear_transport,nessi14_evolution_eqs_pot}, enables to investigate not only the dynamics of the off-equilibrium momentum distribution, but also of two-times correlation functions for large times and system sizes. This allows to envisage the rich relaxation dynamics of nearly-integrable systems, which turns out to be characterized by several relevant timescales that we quantitatively study in the specific example at hand.

We consider a 1D model of spinless fermions with nearest neighbor interactions, and nearest and next-to-nearest neighbor hopping, $H(J_1,J_2,\Delta)=H_0(J_1,J_2)+H_1(\Delta)$,
\begin{equation}\label{eq:ham0} H_0(J_1,J_2)=-J_1\sum_{j=0}^{L-1}(c_{j}^{\dagger}c_{j+1}+\mathrm{h.c.})-J_2\sum_{j=0}^{L-1}(c_{j}^{\dagger}c_{j+2}+\mathrm{h.c.}),
\end{equation}
\begin{equation}\label{eq:ham1}
H_1(\Delta)=\Delta\sum_{j=0}^{L-1}n_{j}n_{j+1},
\end{equation}
where $L$ is the number of sites in the chain, the $c$ operators obey canonical anticommutation relations and $n_j=c^{\dagger}_jc_j$. We assume periodic boundary
conditions, $c^{\#}_{L+m}=c^{\#}_m$. For $J_2=0$ the model is integrable through Bethe ansatz, while for any $J_2\neq0$ the model is non-integrable.

We are
interested in the dynamics of the system starting from a non-equilibrium initial condition $\rho(0)$, $[H,\rho(0)]\neq0$. We shall restrict to the weakly
interacting case $\alpha=\Delta/J_1\ll 1$ and consider only homogeneous initial states.  In such case the operators defining the momentum space density distribution $\hat{n}_k=\frac{1}{L}\sum_{i,j}e^{-\frac{2\pi
k}{L}(i-j)}c^{\dagger}_i c_j$, with $k=0,\ldots,L-1$, emerge as the natural slow variables of the system.
Using the projection operator technique it is possible
to derive an \emph{exact} coupled system of equations of motion for the average $n(k,t)=\langle \hat{n}_k(t)\rangle$, where $\langle\ldots\rangle=\mathrm{Tr}[\rho(0)\ldots]$, and the
fluctuations $F(k,k',t)=\langle \hat{n}_k(t)\hat{n}_{k'}\rangle-\langle \hat{n}_k(t)\rangle\langle\hat{n}_{k'}\rangle$ of the slow variables~\cite{grabert82_book_pot,rau95_review_pot,breuer82_book_open_systems,robertson66_pot_first,kawasaki73_non-linear_transport,nessi14_evolution_eqs_pot}. We work in the
Heisenberg picture, $\hat{\mathcal{O}}(t)=e^{iHt}\hat{\mathcal{O}}e^{-iHt}$ ($\hslash=1$).
The integro-differential equations are manageable if
\textit{(i)} we restrict to uncorrelated initial conditions of the form $\rho(0)=\frac{1}{Z}e^{-\sum_k\lambda(k,0)\hat{n}_k}$, with
$Z=\mathrm{Tr}[e^{-\sum_k\lambda(k,0)\hat{n}_k}]$, which include free fermion initial states, and \textit{(ii)} we keep only
the leading order in $\alpha$. In particular, the equation for the averages $n(k,t)$ is formally identical to that derived using heuristic arguments in Ref.~\cite{stark13_kinetic_description}. For the explicit expressions and a thorough derivation of such evolution equations see~\cite{nessi14_evolution_eqs_pot,supplementary}. Although the approach involves perturbative steps, the results go beyond conventional lowest order perturbation theory because the
perturbation expansion is performed inside the integro-differential equations and, therefore, the coupling is involved in a highly non-linear way in the final
expressions. The equations of motion can be efficiently solved numerically allowing to study large systems ($L\sim10^3$) and times far beyond the reach of
standard numerical techniques such as time-dependent density-matrix renormalization group (t-DMRG), which allows to access the long time dynamics of the system in the thermodynamic limit. We fix $\alpha=0.2$ for the rest of the paper.

\begin{figure}
  \includegraphics[width=\largefigwidth]{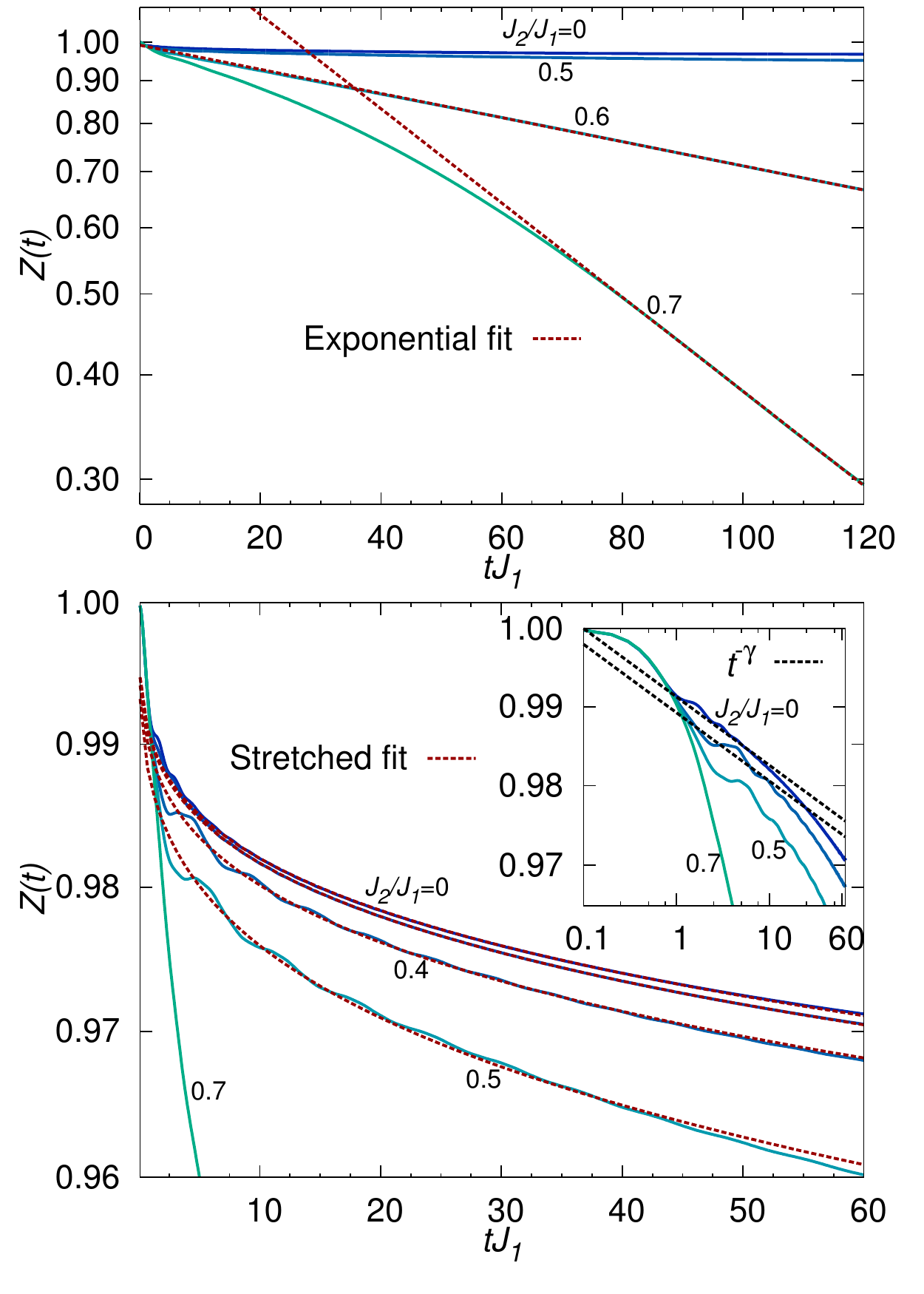}\\
  \caption{Decay of the quasiparticle residue $Z(t)$ varying $J_2$ for a system with $L=1024$ away from any revival. Upper panel: Plot in semilogarithmic scale. Dashed (red) lines are exponential fits. Lower panel: Semi logarithmic plot showing the details of the decay for $J_2\leq0.5J_1$, the curve for $J_2=0.7J_1$ is also presented for comparison. Dashed (red) lines are stretched exponential fits. Inset: Log-log plot. Dashed (black) lines are power laws with the exponent given by the LM predictions $Z(t)\sim t^{-\gamma}$, with $\gamma=\frac{1}{4}(K^{2}+K^{-2}-2)$. The Luttinger parameter is obtained from bosonization, $K=\sqrt{\frac{\pi v_{F}-\Delta}{\pi v_{F}+3\Delta}}$, where $v_F$ is the Fermi velocity.}\label{fig:graphz}
\end{figure}

We begin by considering the effect of varying the strength of the integrability breaking parameter $J_2$. As initial state we consider the ground state $\vert\Psi_0\rangle$ of $H_0(J_1,0)$ with momentum distribution $n(k,0)=\theta(\epsilon_0(k)-\epsilon_0(k_F))$, where $\epsilon_0(k)=-2 J_1 \cos(2k\pi/L)$. We will work at half-filling, $k_F=L/4$. We shall first focus on the relaxation of the momentum modes close to the Fermi surface, whose relaxation timescales are the longest among all momentum modes. In particular, we will study the decay of the quasiparticle residue, defined as $Z(t)=n(k_F,t)-n(k_F+1,t)$. In higher dimensions this quantity exhibits typical ``prethermalization plateaus'', and has become a standard tool to detect metastable states in fermionic systems~\cite{moeckel08_quench_hubbard_high_d,eckstein09_thermalization_quench_fermi_hubbard,nessi14_gge_fermi_liquid}. In $1D$ this quantity does not exhibit plateaus~\cite{hamerla13_1D_quench}. We shall show that, nevertheless, metastable states are present and that they profoundly affect the relaxation of the momentum modes close to the Fermi points. In the upper panel of Figure~\ref{fig:graphz} we show the decay of the quasiparticle residue for different values of $J_2$ for a system with $L=1024$ and for times far away from any recurrence effect. In such situation finite size corrections are negligible and we are thus accessing the thermodynamic limit dynamics. It is clearly visible that for $J_2\leq 0.5 J_1$ the decay of $Z(t)$ is extremely slow (subexponential) while for $J_2> 0.5 J_1$ it exhibits exponential behavior. In the lower panel we analyze in more detail the slow evolution. In particular, we find that after a fast Gaussian-like initial evolution taking place for $tJ_1<1$, there is a faster than power law but slower than exponential decay. In this regime it is possible to fit a stretched exponential $Z(t)=e^{-(t/\tau)^{\beta}}$, where both $\tau$ and $\beta$ depend on $\Delta$ and $J_2$. In the fits showed in the lower panel of Figure~\ref{fig:graphz}, $\tau J_{1}\sim10^{7}$ and $\beta\sim 0.27$. On the other hand, for the exponential decays taking place for $J_2> 0.5 J_1$, $\tau J_{1}\sim10^{2}$. Below we shall see that such abrupt decrease of the relaxation time scales is related to an abrupt increase in the lifetime of the metastable states. We finally note that for $J_2\leq 0.5 J_1$, after the initial Gaussian evolution, the initial trend of the curves is very well described by the power law decay predicted by the non-equilibrium dynamics of the Luttinger model~\cite{cazalilla06_quench_LL,iucci09_quench_LL}. In the t-DMRG study in Ref.~\cite{karrasch12_ll_universality_quench}, where this remarkable fact was first noticed, such initial trend was the only accessible portion of the dynamics. The fact that the evolution equations capture such feature of the dynamics clearly indicates that the results are not perturbative in the usual sense.

\begin{figure}
  \includegraphics[width=\figwidth]{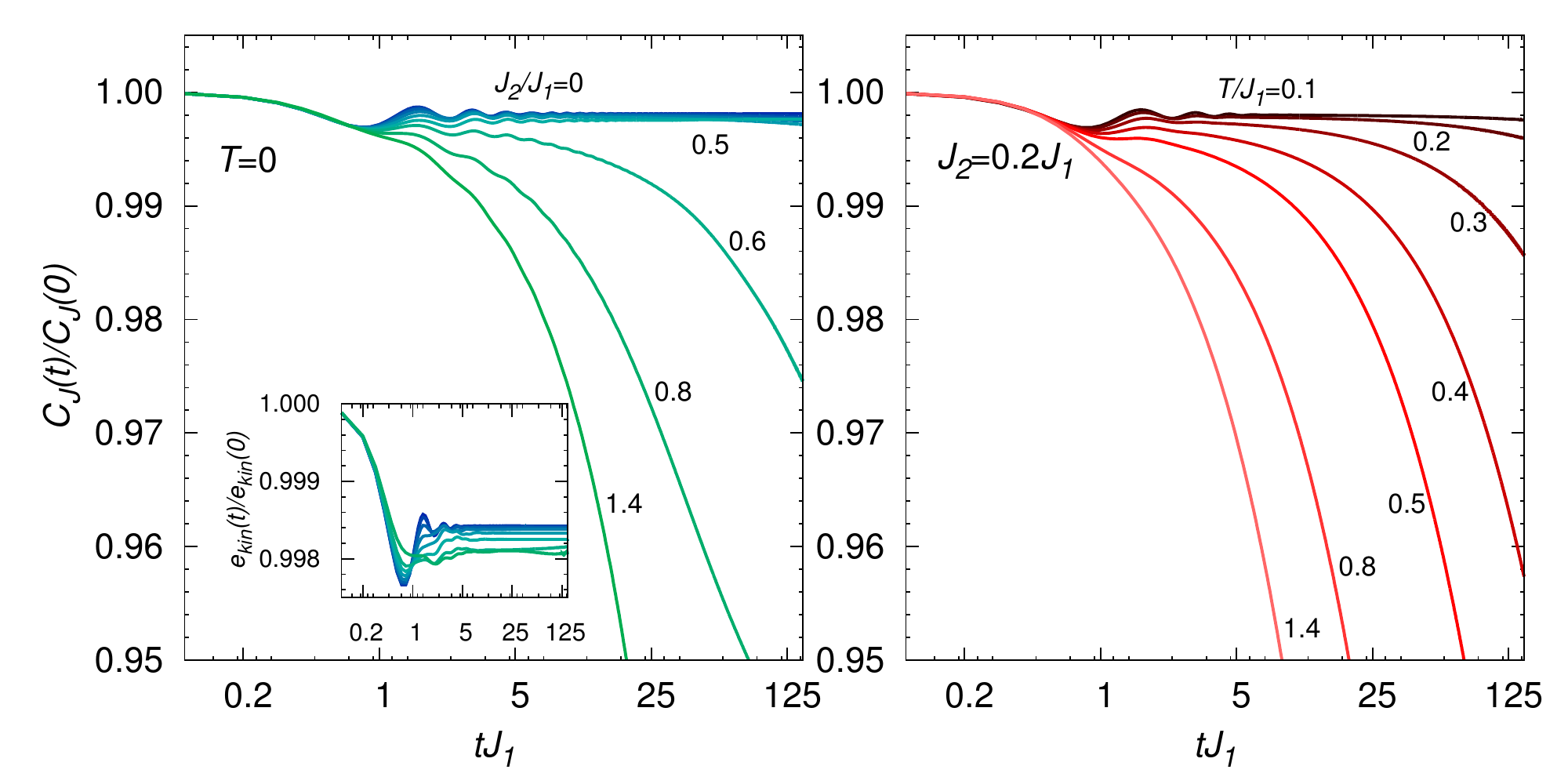}\\
  \caption{Decay of the correlation function $C_J(t)$. Left panel: starting from the $T=0$ initial condition varying $J_2$, for which the energy density is $e=0$ independently of the value of $J_2$. Inset: Relaxation of the kinetic energy for the same parameters as the main figure. Right panel: for fixed $J_2=0.2J_1$ varying the temperature of the initial state. Note the logarithmic scale on the time axis.}\label{fig:graphcjkin}
\end{figure}

Two-time correlations turn out to be an adequate tool to detect and analyze the prethermalized states. In particular, we introduce the connected correlation function of the current operator $\hat{J}=\frac{1}{2i}\sum_{j}c^{\dagger}_{j}c_{j+1}-c^{\dagger}_{j+1}c_{j}$,
\begin{equation}
C_J(t)=\langle \hat{J}(t)\hat{J}\rangle-\langle \hat{J}(t)\rangle\langle\hat{J}\rangle,
\end{equation}
which can be obtained from the fluctuations of the slow variables $C_J(t)=\sum_{k,k'}\sin(2k\pi/L)\sin(2k'\pi/L)F(k,k',t)$. It is important to note that two-times connected correlation functions of local observables are a standard tool to diagnose the ergodic status of \emph{equilibrium} dynamics, both in quantum~\cite{jona-lasiinio96_mixing_quantum_thermo_limit,prosen98_integrability-ergodicity_thermo_limit,prosen99_ergodicity_kicked_xxz} and classical systems~\cite{cavagna09_supercooled_pedestrians}. Loosely speaking, this type of correlators is expected to decay rapidly to zero for ergodic systems, since the initial state and the state at time $t$ are expected to be completely decorrelated after some characteristic timescale, i.e., the system is expected to loose memory of the initial condition. For non-ergodic systems the correlator is expected to saturate to a non-zero constant. An intermediate behavior, with a plateau emerging in between a first fast evolution and the final decay to zero, arises in systems that get caught into long-lived metastable states, the most prominent example being glassy systems, such as spin glasses and supercooled liquids~\cite{castellani05_spin-glass_pedestrians,cavagna09_supercooled_pedestrians}. On general grounds we expect that this kind of correlators should perform as a similar diagnosis tool in the non-equilibrium situation under consideration. In the left panel of Fig.~\ref{fig:graphcjkin} we show the decay of $C_J(t)$ for the ground state initial condition varying the value of the integrability breaking parameter $J_2$ in a system with $L=800$ and for times away from any recurrence effects, which, again, amounts to investigate the dynamics in the thermodynamic limit. For large values $J_2\sim J_1$ the decay develops in a single step. Decreasing $J_2$ a plateau arises in between the initial Gaussian evolution and the final decay. The length of the plateau becomes larger as we further decrease $J_2$ yet it does not increase smoothly but rather seems to have an abrupt jump exactly for $J_2=0.5J_1$. This clearly indicates that the system is caught in non-thermal metastable states whose lifetime grows abruptly around $J_2\sim 0.5J_1$.

Another factor that deeply influences the lifetime of such prethermalized states turns out to be the energy of the system. In particular, we investigated the behavior of $C_J(t)$ starting from finite temperature initial states $n(k,0)=(1+\exp[(\epsilon(k)-\epsilon(k_F))/T])^{-1}$, where $T$ is the temperature ($k_B=1$). The energy density of the system $e=\frac{1}{L}\mathrm{Tr}[\rho(0)H]-e_0$, with $e_0=\frac{1}{L}\langle\Psi_0\vert H(J_1,0,\Delta)\vert\Psi_0\rangle$, is a smooth, monotonous function of $T$. In the right panel of Fig.~\ref{fig:graphcjkin} we show the relaxation of $C_J(t)$ for $J_2=0.2$ varying the temperature of the initial condition. The effect of increasing the temperature (energy) of the initial state is to gradually decrease the lifetime of the prethermalized states. In particular, for sufficiently high energies, they are completely suppressed. We observe that the metastable states emerge in the same timescale in which the kinetic energy of the system $e_{kin}(t)=\langle H_0(t)\rangle$ saturates to its final value (see inset in Fig.~\ref{fig:graphcjkin}). Such prethermalization timescale turns out to be independent of the value of $J_2$ and the energy of the system: $t_{pt}J_1\sim 5$ in all cases.

A simple qualitative picture can be formulated. The initial fast evolution taking place for $tJ_{1}<1$ is caused by the dephasing of some quasifree modes of the system~\footnote{Dephasing dynamics was early studied in connection with the mean field approximation to the non-equilibrium dynamics of isolated systems~\cite{habib96_dephasing_mean_field}. In such case, the quasifree modes are the modes diagonalizing the quadratic mean field Hamiltonian.}. For fermionic systems these can be identified with the bosonic modes associated with the bosonization of the excitations close to the Fermi points surface~\cite{nessi14_gge_fermi_liquid}. This initial regime is not sensitive to the details of the interaction and is analogous to the initial ballistic expansion in the relaxation of classical glasses and fluids in general. The subsequent relaxation of the system is provided by inelastic collisions. If the inelastic relaxation channels are scarce the system becomes trapped in metastable states whose lifetime is a measure of the rate of occurrence of such inelastic scattering events. For the model $H(J_1,J_2,\Delta)$ it can be shown that as soon as $J_2$ becomes larger than $0.5 J_1$ the manifold of kinetically allowed collisions (those that conserve momentum \emph{and} kinetic energy) is dramatically enlarged~\cite{furst13_qbe_hubbard_1d_nnn}. Such type of collisions are included in (but do not exhaust) the equations of motion that we consider~\cite{supplementary}. The number of relaxation channels is thus drastically enlarged beyond a \emph{finite} threshold away from integrability. Moreover, since at higher energies there are more possible inelastic collisions, the lifetime of the metastable states is suppressed as we increase the energy of the system.

\begin{figure}
  \includegraphics[width=\figwidth]{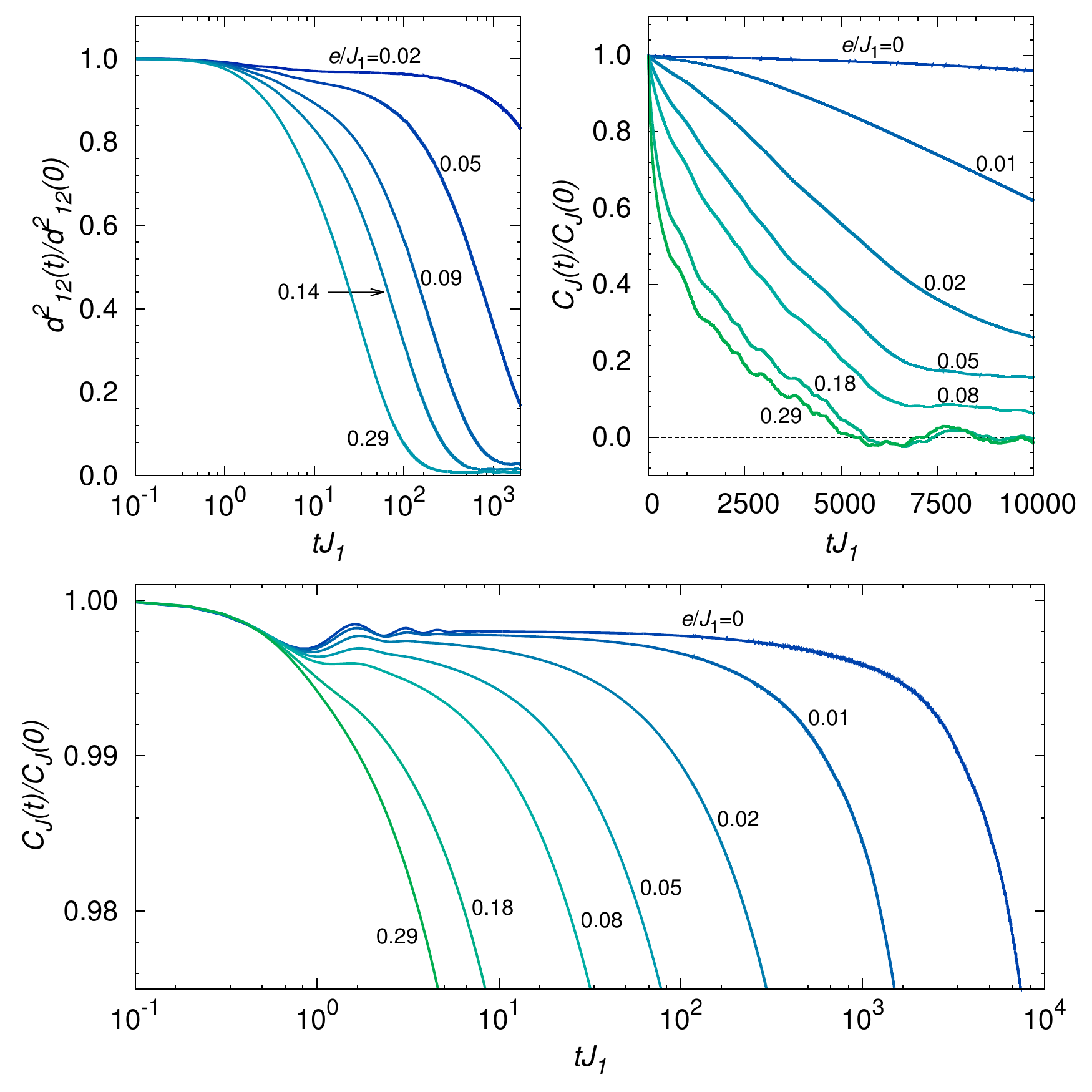}\\
  \caption{Top left panel: Decay of the dynamical distance $d^2_{12}(t)$ (note the logarithmic scale in the time axis) for $J_2=0.2J_1$ varying the energy of the pair of initial conditions. Top right: Overall decay of the correlation function $C_J(t)$ for $J_2=0.2J_1$ varying the temperature $T$ of the initial state. Bottom panel: Detail of the behavior for $C_{J}(t)/C_{J}(0)\sim1$ making visible the presence of two-step relaxation and long-lived quasi-stationary states (note the logarithmic scale on the time axis).}\label{fig:graphcj_rep}
\end{figure}

\begin{figure}
  \includegraphics[width=\largefigwidth]{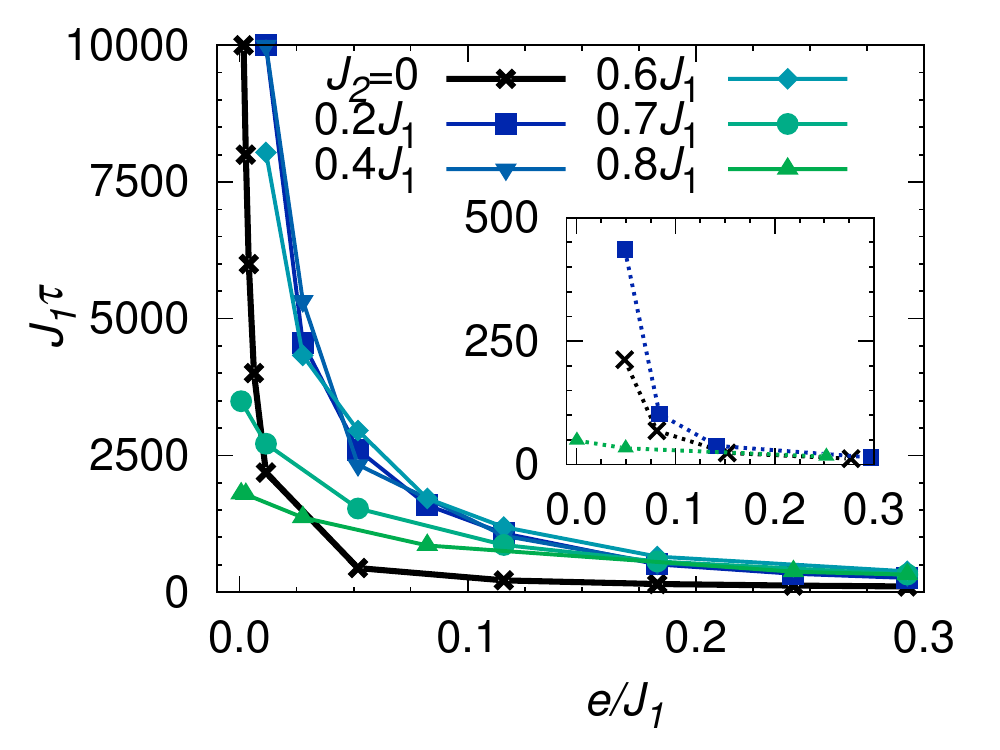}\\
  \caption{Characteristic decay timescales of the correlator $C_J(t)$ for three different representative values of $J_2$ as a function of the energy of the initial state. Inset:  Decay timescales of the dynamical distance $d_{12}$ as a function of the energy density of the pair of initial conditions for three representative values of $J_2$, $0$, $0.2J_1$ and $0.8J_1$ (with the same symbols as the main figure but with dashed lines).}\label{fig:timescales}
\end{figure}

In order to make quantitative statements about the lifetime of the prethermalized states we find convenient to concentrate on smaller systems, with $L=256$, which shall allow us to access longer times. In this case it must be noted that, specially for long times, deviations from the thermodynamic limit dynamics may be appreciable~\cite{supplementary}. To characterize the overall relaxation of the momentum distribution we study the dynamics of two replicas of the system. In particular, we prepare two initial conditions $n_1(k,0)$ and $n_2(k,0)$ with approximately the same energy and particle density, and define the dynamical distance between the time evolved momentum distributions
\begin{equation}
d^2_{12}(t)=\frac{1}{L}\sum_{k} (n_1(k,t)-n_2(k,t))^2.
\end{equation}
For a system whose only conserved quantities are total energy and particle number this distance should decay to zero at long times. For systems with additional conservation laws (like the integrable model $J_2=0$) it may saturate to a non-zero constant. In any case, the presence of short time plateaus in the time evolution of $d^2_{12}(t)$ represents a clear sign of the formation of metastable states. In other words, if the two systems, prepared in different initial conditions, get caught in different metastable states that are at a distance $d^2_{12,plat}$  of each other, then a plateau in $d^2_{12}(t)$ at the value $d^2_{12,plat}$ would be present. If $d^2_{12,plat}\neq 0$ we can be sure that these are non-thermal metastable states, at least for the non-integrable model. We prepare the initial conditions slightly perturbing free fermion thermal states. In the top left panel of Fig.~\ref{fig:graphcj_rep} we show the decay of $d^2_{12}(t)$ for $L=256$ and $J_2=0.2J_1$, but the results are similar for any $J_2<0.5J_1$~\footnote{In the particular case of $J_2=0$ we also find short time plateaus, but the distance saturate to a non-zero constant.}. We find that if the energy of the pair of initial conditions is low enough $d^2_{12}(t)$ relaxes in two steps. The length of the plateau increases for lower energies. In contrast, for $J_2>0.5J_1$ we find single-step relaxation in all the energy range (not shown), confirming the picture that emerged from the analysis of $C_J(t)$. In the top right panel of Fig.~\ref{fig:graphcj_rep} we show the overall relaxation of the correlation function $C_J(t)$ also for $J_2=0.2J_1$. For high energies we find that it decays to zero on the accessible timescales. For low energies we observe a very pronounced slowing down of the relaxation. In the bottom panel we show that the slowing down is caused by the presence of long-lived metastable states whose large lifetime can be appreciated.

In Fig.~\ref{fig:timescales} we  show the decay timescales of $C_J(t)$ for different values of $J_2$. The relaxation timescale $\tau$ was defined as $C_J(\tau)/C_J(0)=0.6$, in order to extract the maximum possible number of data points from the results at disposal while still considering a faithful indicator of the relaxation timescale. We see that for $J_2\lesssim0.5J_1$ the relaxation timescales show an abrupt increase for sufficiently low energies. 
 For $J_2>0.5J_1$ the relaxation timescale is almost unchanged as we vary the energy of the initial condition.

A special remark is in order with respect to the effective decay timescales of $d^2_{12}(t)$. Being quite independent of the specific form of the initial conditions, this is one of the characteristic timescales in the thermalization process~\cite{berges04_prethermalization_idea}.  It is the timescale beyond which the system has lost all memory of the initial conditions at the level of single particle observables. Nevertheless this does not mean that the system is in equilibrium. In fact, we find that $\tau[C_J]$ is, at least, one order of magnitude larger than $\tau[d_{12}]$ (defined in the same way as for $C_J$) in all cases in which we have data to compare. This is illustrated in the inset of Fig.~\ref{fig:timescales}. The fact that one-particle observables attain thermal behavior much before than two-particle correlations indicates that thermalization is a hierarchical process. The information about the initial conditions encoded in a correlation function increases with its order. Our results suggest that, accordingly, the relaxation timescales also increase monotonically with the order of the correlation function. However, we may also expect that for some finite (but possibly very large) order correlations do not thermalize at all, reflecting the unitarity of quantum dynamics. Finally, we note that it is for times larger than $\tau[d_{12}]$ that a description based on kinetic (memoryless) equations, such as the quantum Boltzmann equation~\cite{erdos04_qbe}, is justified~\cite{berges06_validity_transp_eqs}.

\acknowledgments

This work was partially supported by CONICET (PIP 0662), ANPCyT (PICT 2010-1907) and UNLP (PID X497), Argentina.


\begin{thebibliography}{42}%
\makeatletter
\providecommand \@ifxundefined [1]{%
 \@ifx{#1\undefined}
}%
\providecommand \@ifnum [1]{%
 \ifnum #1\expandafter \@firstoftwo
 \else \expandafter \@secondoftwo
 \fi
}%
\providecommand \@ifx [1]{%
 \ifx #1\expandafter \@firstoftwo
 \else \expandafter \@secondoftwo
 \fi
}%
\providecommand \natexlab [1]{#1}%
\providecommand \enquote  [1]{``#1''}%
\providecommand \bibnamefont  [1]{#1}%
\providecommand \bibfnamefont [1]{#1}%
\providecommand \citenamefont [1]{#1}%
\providecommand \href@noop [0]{\@secondoftwo}%
\providecommand \href [0]{\begingroup \@sanitize@url \@href}%
\providecommand \@href[1]{\@@startlink{#1}\@@href}%
\providecommand \@@href[1]{\endgroup#1\@@endlink}%
\providecommand \@sanitize@url [0]{\catcode `\\12\catcode `\$12\catcode
  `\&12\catcode `\#12\catcode `\^12\catcode `\_12\catcode `\%12\relax}%
\providecommand \@@startlink[1]{}%
\providecommand \@@endlink[0]{}%
\providecommand \url  [0]{\begingroup\@sanitize@url \@url }%
\providecommand \@url [1]{\endgroup\@href {#1}{\urlprefix }}%
\providecommand \urlprefix  [0]{URL }%
\providecommand \Eprint [0]{\href }%
\providecommand \doibase [0]{http://dx.doi.org/}%
\providecommand \selectlanguage [0]{\@gobble}%
\providecommand \bibinfo  [0]{\@secondoftwo}%
\providecommand \bibfield  [0]{\@secondoftwo}%
\providecommand \translation [1]{[#1]}%
\providecommand \BibitemOpen [0]{}%
\providecommand \bibitemStop [0]{}%
\providecommand \bibitemNoStop [0]{.\EOS\space}%
\providecommand \EOS [0]{\spacefactor3000\relax}%
\providecommand \BibitemShut  [1]{\csname bibitem#1\endcsname}%
\let\auto@bib@innerbib\@empty
\bibitem [{\citenamefont {Rigol}\ \emph {et~al.}(2008)\citenamefont {Rigol},
  \citenamefont {Dunjko},\ and\ \citenamefont
  {Olshanii}}]{rigol08_mechanism_thermalization}%
  \BibitemOpen
  \bibfield  {author} {\bibinfo {author} {\bibfnamefont {M.}~\bibnamefont
  {Rigol}}, \bibinfo {author} {\bibfnamefont {V.}~\bibnamefont {Dunjko}}, \
  and\ \bibinfo {author} {\bibfnamefont {M.}~\bibnamefont {Olshanii}},\ }\href
  {\doibase doi:10.1038/nature06838} {\bibfield  {journal} {\bibinfo  {journal}
  {Nature (London)}\ }\textbf {\bibinfo {volume} {452}},\ \bibinfo {pages}
  {854} (\bibinfo {year} {2008})}\BibitemShut {NoStop}%
\bibitem [{\citenamefont {Rigol}(2014)}]{rigol14_quenches_thermo_limit}%
  \BibitemOpen
  \bibfield  {author} {\bibinfo {author} {\bibfnamefont {M.}~\bibnamefont
  {Rigol}},\ }\href@noop {} {\bibfield  {journal} {\bibinfo  {journal} {Phys.
  Rev. Lett.}\ }\textbf {\bibinfo {volume} {112}},\ \bibinfo {pages} {170601}
  (\bibinfo {year} {2014})}\BibitemShut {NoStop}%
\bibitem [{\citenamefont {Santos}\ and\ \citenamefont
  {Rigol}(2010)}]{santos10_quantum_chaos_1D}%
  \BibitemOpen
  \bibfield  {author} {\bibinfo {author} {\bibfnamefont {L.~F.}\ \bibnamefont
  {Santos}}\ and\ \bibinfo {author} {\bibfnamefont {M.}~\bibnamefont {Rigol}},\
  }\href@noop {} {\bibfield  {journal} {\bibinfo  {journal} {Phys. Rev. E}\
  }\textbf {\bibinfo {volume} {81}},\ \bibinfo {pages} {036206} (\bibinfo
  {year} {2010})}\BibitemShut {NoStop}%
\bibitem [{\citenamefont {Moeckel}\ and\ \citenamefont
  {Kehrein}(2008)}]{moeckel08_quench_hubbard_high_d}%
  \BibitemOpen
  \bibfield  {author} {\bibinfo {author} {\bibfnamefont {M.}~\bibnamefont
  {Moeckel}}\ and\ \bibinfo {author} {\bibfnamefont {S.}~\bibnamefont
  {Kehrein}},\ }\href {\doibase 10.1103/PhysRevLett.100.175702} {\bibfield
  {journal} {\bibinfo  {journal} {Phys. Rev. Lett.}\ }\textbf {\bibinfo
  {volume} {100}},\ \bibinfo {pages} {175702} (\bibinfo {year}
  {2008})}\BibitemShut {NoStop}%
\bibitem [{\citenamefont {Eckstein}\ \emph {et~al.}(2009)\citenamefont
  {Eckstein}, \citenamefont {Kollar},\ and\ \citenamefont
  {Werner}}]{eckstein09_thermalization_quench_fermi_hubbard}%
  \BibitemOpen
  \bibfield  {author} {\bibinfo {author} {\bibfnamefont {M.}~\bibnamefont
  {Eckstein}}, \bibinfo {author} {\bibfnamefont {M.}~\bibnamefont {Kollar}}, \
  and\ \bibinfo {author} {\bibfnamefont {P.}~\bibnamefont {Werner}},\ }\href
  {\doibase 10.1103/PhysRevLett.103.056403} {\bibfield  {journal} {\bibinfo
  {journal} {Phys. Rev. Lett.}\ }\textbf {\bibinfo {volume} {103}},\ \bibinfo
  {pages} {056403} (\bibinfo {year} {2009})}\BibitemShut {NoStop}%
\bibitem [{\citenamefont {Marino}\ and\ \citenamefont
  {Silva}(2012)}]{marino12_ising_noisy}%
  \BibitemOpen
  \bibfield  {author} {\bibinfo {author} {\bibfnamefont {J.}~\bibnamefont
  {Marino}}\ and\ \bibinfo {author} {\bibfnamefont {A.}~\bibnamefont {Silva}},\
  }\href {\doibase 10.1103/PhysRevB.86.060408} {\bibfield  {journal} {\bibinfo
  {journal} {Phys. Rev. B}\ }\textbf {\bibinfo {volume} {86}},\ \bibinfo
  {pages} {060408} (\bibinfo {year} {2012})}\BibitemShut {NoStop}%
\bibitem [{\citenamefont {van~den Worm}\ \emph {et~al.}(2013)\citenamefont
  {van~den Worm}, \citenamefont {Sawyer}, \citenamefont {Bollinger},\ and\
  \citenamefont {Kastner}}]{vandenworm_2013}%
  \BibitemOpen
  \bibfield  {author} {\bibinfo {author} {\bibfnamefont {M.}~\bibnamefont
  {van~den Worm}}, \bibinfo {author} {\bibfnamefont {B.}~\bibnamefont
  {Sawyer}}, \bibinfo {author} {\bibfnamefont {J.}~\bibnamefont {Bollinger}}, \
  and\ \bibinfo {author} {\bibfnamefont {M.}~\bibnamefont {Kastner}},\
  }\href@noop {} {\bibfield  {journal} {\bibinfo  {journal} {New J. of Phys}\
  }\textbf {\bibinfo {volume} {15}},\ \bibinfo {pages} {83007} (\bibinfo {year}
  {2013})}\BibitemShut {NoStop}%
\bibitem [{\citenamefont {Essler}\ \emph {et~al.}(2014)\citenamefont {Essler},
  \citenamefont {Kehrein}, \citenamefont {Manmana},\ and\ \citenamefont
  {Robinson}}]{essler13_quench_tuneable_integrability_breaking}%
  \BibitemOpen
  \bibfield  {author} {\bibinfo {author} {\bibfnamefont {F.~H.~L.}\
  \bibnamefont {Essler}}, \bibinfo {author} {\bibfnamefont {S.}~\bibnamefont
  {Kehrein}}, \bibinfo {author} {\bibfnamefont {S.~R.}\ \bibnamefont
  {Manmana}}, \ and\ \bibinfo {author} {\bibfnamefont {N.~J.}\ \bibnamefont
  {Robinson}},\ }\href {\doibase 10.1103/PhysRevB.89.165104} {\bibfield
  {journal} {\bibinfo  {journal} {Phys. Rev. B}\ }\textbf {\bibinfo {volume}
  {89}},\ \bibinfo {pages} {165104} (\bibinfo {year} {2014})}\BibitemShut
  {NoStop}%
\bibitem [{\citenamefont {Kollar}\ \emph {et~al.}(2011)\citenamefont {Kollar},
  \citenamefont {Wolf},\ and\ \citenamefont
  {Eckstein}}]{kollar11_gge_pretherm}%
  \BibitemOpen
  \bibfield  {author} {\bibinfo {author} {\bibfnamefont {M.}~\bibnamefont
  {Kollar}}, \bibinfo {author} {\bibfnamefont {F.~A.}\ \bibnamefont {Wolf}}, \
  and\ \bibinfo {author} {\bibfnamefont {M.}~\bibnamefont {Eckstein}},\ }\href
  {\doibase 10.1103/PhysRevB.84.054304} {\bibfield  {journal} {\bibinfo
  {journal} {Phys. Rev. B}\ }\textbf {\bibinfo {volume} {84}},\ \bibinfo
  {pages} {054304} (\bibinfo {year} {2011})}\BibitemShut {NoStop}%
\bibitem [{\citenamefont {Stark}\ and\ \citenamefont
  {Kollar}()}]{stark13_kinetic_description}%
  \BibitemOpen
  \bibfield  {author} {\bibinfo {author} {\bibfnamefont {M.}~\bibnamefont
  {Stark}}\ and\ \bibinfo {author} {\bibfnamefont {M.}~\bibnamefont {Kollar}},\
  }\href {http://arxiv.org/abs/1308.1610} {\ }\Eprint
  {http://arxiv.org/abs/arXiv:1308.1610} {arXiv:1308.1610} \BibitemShut
  {NoStop}%
\bibitem [{\citenamefont {Nessi}\ \emph {et~al.}(2014)\citenamefont {Nessi},
  \citenamefont {Iucci},\ and\ \citenamefont
  {Cazalilla}}]{nessi14_gge_fermi_liquid}%
  \BibitemOpen
  \bibfield  {author} {\bibinfo {author} {\bibfnamefont {N.}~\bibnamefont
  {Nessi}}, \bibinfo {author} {\bibfnamefont {A.}~\bibnamefont {Iucci}}, \ and\
  \bibinfo {author} {\bibfnamefont {M.~A.}\ \bibnamefont {Cazalilla}},\
  }\href@noop {} {\bibfield  {journal} {\bibinfo  {journal} {Phys. Rev. Lett.}\
  }\textbf {\bibinfo {volume} {113}},\ \bibinfo {pages} {210402} (\bibinfo
  {year} {2014})}\BibitemShut {NoStop}%
\bibitem [{\citenamefont {Marcuzzi}\ \emph {et~al.}(2013)\citenamefont
  {Marcuzzi}, \citenamefont {Marino}, \citenamefont {Gambassi},\ and\
  \citenamefont {Silva}}]{marcuzzi13_pretherm_ising}%
  \BibitemOpen
  \bibfield  {author} {\bibinfo {author} {\bibfnamefont {M.}~\bibnamefont
  {Marcuzzi}}, \bibinfo {author} {\bibfnamefont {J.}~\bibnamefont {Marino}},
  \bibinfo {author} {\bibfnamefont {A.}~\bibnamefont {Gambassi}}, \ and\
  \bibinfo {author} {\bibfnamefont {A.}~\bibnamefont {Silva}},\ }\href@noop {}
  {\bibfield  {journal} {\bibinfo  {journal} {Phys. Rev. Lett.}\ }\textbf
  {\bibinfo {volume} {111}},\ \bibinfo {pages} {197203} (\bibinfo {year}
  {2013})}\BibitemShut {NoStop}%
\bibitem [{\citenamefont {Mitra}(2013)}]{mitra_prethermalization_2013}%
  \BibitemOpen
  \bibfield  {author} {\bibinfo {author} {\bibfnamefont {A.}~\bibnamefont
  {Mitra}},\ }\href {\doibase 10.1103/PhysRevB.87.205109} {\bibfield  {journal}
  {\bibinfo  {journal} {Phys. Rev. B}\ }\textbf {\bibinfo {volume} {87}},\
  \bibinfo {pages} {205109} (\bibinfo {year} {2013})}\BibitemShut {NoStop}%
\bibitem [{\citenamefont {Gring}\ \emph {et~al.}(2012)\citenamefont {Gring},
  \citenamefont {Kuhnert}, \citenamefont {Langen}, \citenamefont {Kitagawa},
  \citenamefont {Rauer}, \citenamefont {Schreitl}, \citenamefont {Mazets},
  \citenamefont {Smith}, \citenamefont {Demler},\ and\ \citenamefont
  {Schmiedmayer}}]{gring12_prethermalization_isolated_bose_gas}%
  \BibitemOpen
  \bibfield  {author} {\bibinfo {author} {\bibfnamefont {M.}~\bibnamefont
  {Gring}}, \bibinfo {author} {\bibfnamefont {M.}~\bibnamefont {Kuhnert}},
  \bibinfo {author} {\bibfnamefont {T.}~\bibnamefont {Langen}}, \bibinfo
  {author} {\bibfnamefont {T.}~\bibnamefont {Kitagawa}}, \bibinfo {author}
  {\bibfnamefont {B.}~\bibnamefont {Rauer}}, \bibinfo {author} {\bibfnamefont
  {M.}~\bibnamefont {Schreitl}}, \bibinfo {author} {\bibfnamefont
  {I.}~\bibnamefont {Mazets}}, \bibinfo {author} {\bibfnamefont {D.~A.}\
  \bibnamefont {Smith}}, \bibinfo {author} {\bibfnamefont {E.}~\bibnamefont
  {Demler}}, \ and\ \bibinfo {author} {\bibfnamefont {J.}~\bibnamefont
  {Schmiedmayer}},\ }\href {10.1126/science.1224953} {\bibfield  {journal}
  {\bibinfo  {journal} {Science}\ }\textbf {\bibinfo {volume} {337}},\ \bibinfo
  {pages} {1318} (\bibinfo {year} {2012})}\BibitemShut {NoStop}%
\bibitem [{\citenamefont {Langen}\ \emph {et~al.}(2015)\citenamefont {Langen},
  \citenamefont {Erne}, \citenamefont {Geiger}, \citenamefont {Rauer},
  \citenamefont {Schweigler}, \citenamefont {Kuhnert}, \citenamefont
  {Rohringer}, \citenamefont {Mazets}, \citenamefont {Gasenzer},\ and\
  \citenamefont {Schmiedmayer}}]{langen15_quench_lieb-liniger}%
  \BibitemOpen
  \bibfield  {author} {\bibinfo {author} {\bibfnamefont {T.}~\bibnamefont
  {Langen}}, \bibinfo {author} {\bibfnamefont {S.}~\bibnamefont {Erne}},
  \bibinfo {author} {\bibfnamefont {R.}~\bibnamefont {Geiger}}, \bibinfo
  {author} {\bibfnamefont {B.}~\bibnamefont {Rauer}}, \bibinfo {author}
  {\bibfnamefont {T.}~\bibnamefont {Schweigler}}, \bibinfo {author}
  {\bibfnamefont {M.}~\bibnamefont {Kuhnert}}, \bibinfo {author} {\bibfnamefont
  {W.}~\bibnamefont {Rohringer}}, \bibinfo {author} {\bibfnamefont {I.~E.}\
  \bibnamefont {Mazets}}, \bibinfo {author} {\bibfnamefont {T.}~\bibnamefont
  {Gasenzer}}, \ and\ \bibinfo {author} {\bibfnamefont {J.}~\bibnamefont
  {Schmiedmayer}},\ }\href@noop {} {\bibfield  {journal} {\bibinfo  {journal}
  {unpublished}\ } (\bibinfo {year} {2015})},\ \Eprint
  {http://arxiv.org/abs/arXiv:1411.7185} {arXiv:1411.7185} \BibitemShut
  {NoStop}%
\bibitem [{\citenamefont {Berges}\ \emph {et~al.}(2004)\citenamefont {Berges},
  \citenamefont {Bors\'anyi},\ and\ \citenamefont
  {Wetterich}}]{berges04_prethermalization_idea}%
  \BibitemOpen
  \bibfield  {author} {\bibinfo {author} {\bibfnamefont {J.}~\bibnamefont
  {Berges}}, \bibinfo {author} {\bibfnamefont {S.}~\bibnamefont {Bors\'anyi}},
  \ and\ \bibinfo {author} {\bibfnamefont {C.}~\bibnamefont {Wetterich}},\
  }\href {\doibase 10.1103/PhysRevLett.93.142002} {\bibfield  {journal}
  {\bibinfo  {journal} {Phys. Rev. Lett.}\ }\textbf {\bibinfo {volume} {93}},\
  \bibinfo {pages} {142002} (\bibinfo {year} {2004})}\BibitemShut {NoStop}%
\bibitem [{\citenamefont {Poletti}\ \emph {et~al.}(2013)\citenamefont
  {Poletti}, \citenamefont {Barmettler}, \citenamefont {Georges},\ and\
  \citenamefont {Kollath}}]{poletti13_glass_open_hubbard}%
  \BibitemOpen
  \bibfield  {author} {\bibinfo {author} {\bibfnamefont {D.}~\bibnamefont
  {Poletti}}, \bibinfo {author} {\bibfnamefont {P.}~\bibnamefont {Barmettler}},
  \bibinfo {author} {\bibfnamefont {A.}~\bibnamefont {Georges}}, \ and\
  \bibinfo {author} {\bibfnamefont {C.}~\bibnamefont {Kollath}},\ }\href@noop
  {} {\bibfield  {journal} {\bibinfo  {journal} {Phys. Rev. Lett.}\ }\textbf
  {\bibinfo {volume} {111}},\ \bibinfo {pages} {195301} (\bibinfo {year}
  {2013})}\BibitemShut {NoStop}%
\bibitem [{\citenamefont {Sciolla}\ \emph {et~al.}(2014)\citenamefont
  {Sciolla}, \citenamefont {Poletti},\ and\ \citenamefont
  {Kollath}}]{sciolla14_ageing_open_hubbard}%
  \BibitemOpen
  \bibfield  {author} {\bibinfo {author} {\bibfnamefont {B.}~\bibnamefont
  {Sciolla}}, \bibinfo {author} {\bibfnamefont {D.}~\bibnamefont {Poletti}}, \
  and\ \bibinfo {author} {\bibfnamefont {C.}~\bibnamefont {Kollath}},\
  }\href@noop {} {\bibfield  {journal} {\bibinfo  {journal} {unpublished}\ }
  (\bibinfo {year} {2014})},\ \Eprint {http://arxiv.org/abs/arXiv:1407.4939}
  {arXiv:1407.4939} \BibitemShut {NoStop}%
\bibitem [{\citenamefont {Grabert}(1982)}]{grabert82_book_pot}%
  \BibitemOpen
  \bibfield  {author} {\bibinfo {author} {\bibfnamefont {H.}~\bibnamefont
  {Grabert}},\ }\href@noop {} {\emph {\bibinfo {title} {Projection operator
  techniques in nonequilibrium statistical mechanics}}}\ (\bibinfo  {publisher}
  {Springer-Verlags},\ \bibinfo {year} {1982})\BibitemShut {NoStop}%
\bibitem [{\citenamefont {Rau}\ and\ \citenamefont
  {M\"{u}ller}(1996)}]{rau95_review_pot}%
  \BibitemOpen
  \bibfield  {author} {\bibinfo {author} {\bibfnamefont {J.}~\bibnamefont
  {Rau}}\ and\ \bibinfo {author} {\bibfnamefont {B.}~\bibnamefont
  {M\"{u}ller}},\ }\href@noop {} {\bibfield  {journal} {\bibinfo  {journal}
  {Phys. Rep.}\ }\textbf {\bibinfo {volume} {272}},\ \bibinfo {pages} {1}
  (\bibinfo {year} {1996})}\BibitemShut {NoStop}%
\bibitem [{\citenamefont {Breuer}\ and\ \citenamefont
  {Petruccione}(2002)}]{breuer82_book_open_systems}%
  \BibitemOpen
  \bibfield  {author} {\bibinfo {author} {\bibfnamefont {H.-P.}\ \bibnamefont
  {Breuer}}\ and\ \bibinfo {author} {\bibfnamefont {F.}~\bibnamefont
  {Petruccione}},\ }\href@noop {} {\emph {\bibinfo {title} {The Theory of Open
  Quantum Systems}}}\ (\bibinfo  {publisher} {Oxford University Press},\
  \bibinfo {year} {2002})\BibitemShut {NoStop}%
\bibitem [{\citenamefont {Robertson}(1966)}]{robertson66_pot_first}%
  \BibitemOpen
  \bibfield  {author} {\bibinfo {author} {\bibfnamefont {B.}~\bibnamefont
  {Robertson}},\ }\href@noop {} {\bibfield  {journal} {\bibinfo  {journal}
  {Phys. Rev.}\ }\textbf {\bibinfo {volume} {144}},\ \bibinfo {pages} {151}
  (\bibinfo {year} {1966})}\BibitemShut {NoStop}%
\bibitem [{\citenamefont {Kawasaki}\ and\ \citenamefont
  {Gunton}(1973)}]{kawasaki73_non-linear_transport}%
  \BibitemOpen
  \bibfield  {author} {\bibinfo {author} {\bibfnamefont {K.}~\bibnamefont
  {Kawasaki}}\ and\ \bibinfo {author} {\bibfnamefont {J.}~\bibnamefont
  {Gunton}},\ }\href@noop {} {\bibfield  {journal} {\bibinfo  {journal} {Phys.
  Rev. A}\ }\textbf {\bibinfo {volume} {8}},\ \bibinfo {pages} {2048} (\bibinfo
  {year} {1973})}\BibitemShut {NoStop}%
\bibitem [{\citenamefont {Nessi}\ and\ \citenamefont
  {Iucci}(2014)}]{nessi14_evolution_eqs_pot}%
  \BibitemOpen
  \bibfield  {author} {\bibinfo {author} {\bibfnamefont {N.}~\bibnamefont
  {Nessi}}\ and\ \bibinfo {author} {\bibfnamefont {A.}~\bibnamefont {Iucci}},\
  }\href@noop {} {\bibfield  {journal} {\bibinfo  {journal} {Jour. Phys.: Conf.
  Ser.}\ }\textbf {\bibinfo {volume} {568}},\ \bibinfo {pages} {012013}
  (\bibinfo {year} {2014})}\BibitemShut {NoStop}%
\bibitem [{sup()}]{supplementary}%
  \BibitemOpen
  \href@noop {} {\ }\bibinfo {note} {See supplementary material for
  details}\BibitemShut {NoStop}%
\bibitem [{\citenamefont {Hamerla}\ and\ \citenamefont
  {Uhrig}(2013)}]{hamerla13_1D_quench}%
  \BibitemOpen
  \bibfield  {author} {\bibinfo {author} {\bibfnamefont {S.~A.}\ \bibnamefont
  {Hamerla}}\ and\ \bibinfo {author} {\bibfnamefont {G.~S.}\ \bibnamefont
  {Uhrig}},\ }\href {http://stacks.iop.org/1367-2630/15/i=7/a=073012}
  {\bibfield  {journal} {\bibinfo  {journal} {New J. Phys.}\ }\textbf {\bibinfo
  {volume} {15}},\ \bibinfo {pages} {073012} (\bibinfo {year}
  {2013})}\BibitemShut {NoStop}%
\bibitem [{\citenamefont {Cazalilla}(2006)}]{cazalilla06_quench_LL}%
  \BibitemOpen
  \bibfield  {author} {\bibinfo {author} {\bibfnamefont {M.~A.}\ \bibnamefont
  {Cazalilla}},\ }\href {\doibase 10.1103/PhysRevLett.97.156403} {\bibfield
  {journal} {\bibinfo  {journal} {Phys. Rev. Lett.}\ }\textbf {\bibinfo
  {volume} {97}},\ \bibinfo {pages} {156403} (\bibinfo {year}
  {2006})}\BibitemShut {NoStop}%
\bibitem [{\citenamefont {Iucci}\ and\ \citenamefont
  {Cazalilla}(2009)}]{iucci09_quench_LL}%
  \BibitemOpen
  \bibfield  {author} {\bibinfo {author} {\bibfnamefont {A.}~\bibnamefont
  {Iucci}}\ and\ \bibinfo {author} {\bibfnamefont {M.~A.}\ \bibnamefont
  {Cazalilla}},\ }\href {\doibase 10.1103/PhysRevA.80.063619} {\bibfield
  {journal} {\bibinfo  {journal} {Phys. Rev. A}\ }\textbf {\bibinfo {volume}
  {80}},\ \bibinfo {pages} {063619} (\bibinfo {year} {2009})}\BibitemShut
  {NoStop}%
\bibitem [{\citenamefont {Karrasch}\ \emph {et~al.}(2012)\citenamefont
  {Karrasch}, \citenamefont {Rentrop}, \citenamefont {Schuricht},\ and\
  \citenamefont {Meden}}]{karrasch12_ll_universality_quench}%
  \BibitemOpen
  \bibfield  {author} {\bibinfo {author} {\bibfnamefont {C.}~\bibnamefont
  {Karrasch}}, \bibinfo {author} {\bibfnamefont {J.}~\bibnamefont {Rentrop}},
  \bibinfo {author} {\bibfnamefont {D.}~\bibnamefont {Schuricht}}, \ and\
  \bibinfo {author} {\bibfnamefont {V.}~\bibnamefont {Meden}},\ }\href
  {\doibase karrasch12_ll_universality_quench} {\bibfield  {journal} {\bibinfo
  {journal} {Phys. Rev. Lett.}\ }\textbf {\bibinfo {volume} {109}},\ \bibinfo
  {pages} {126406} (\bibinfo {year} {2012})}\BibitemShut {NoStop}%
\bibitem [{\citenamefont {Jona-Lasinio}\ and\ \citenamefont
  {Presilla}(1996)}]{jona-lasiinio96_mixing_quantum_thermo_limit}%
  \BibitemOpen
  \bibfield  {author} {\bibinfo {author} {\bibfnamefont {G.}~\bibnamefont
  {Jona-Lasinio}}\ and\ \bibinfo {author} {\bibfnamefont {C.}~\bibnamefont
  {Presilla}},\ }\href@noop {} {\bibfield  {journal} {\bibinfo  {journal}
  {Phys. Rev. Lett.}\ }\textbf {\bibinfo {volume} {77}},\ \bibinfo {pages}
  {4322} (\bibinfo {year} {1996})}\BibitemShut {NoStop}%
\bibitem [{\citenamefont
  {Prosen}(1998)}]{prosen98_integrability-ergodicity_thermo_limit}%
  \BibitemOpen
  \bibfield  {author} {\bibinfo {author} {\bibfnamefont {T.}~\bibnamefont
  {Prosen}},\ }\href@noop {} {\bibfield  {journal} {\bibinfo  {journal} {Phys.
  Rev. Lett.}\ }\textbf {\bibinfo {volume} {80}},\ \bibinfo {pages} {1808}
  (\bibinfo {year} {1998})}\BibitemShut {NoStop}%
\bibitem [{\citenamefont {Prosen}(1999)}]{prosen99_ergodicity_kicked_xxz}%
  \BibitemOpen
  \bibfield  {author} {\bibinfo {author} {\bibfnamefont {T.}~\bibnamefont
  {Prosen}},\ }\href@noop {} {\bibfield  {journal} {\bibinfo  {journal} {Phys.
  Rev. E}\ }\textbf {\bibinfo {volume} {60}},\ \bibinfo {pages} {3949}
  (\bibinfo {year} {1999})}\BibitemShut {NoStop}%
\bibitem [{\citenamefont {Cavagna}(2009)}]{cavagna09_supercooled_pedestrians}%
  \BibitemOpen
  \bibfield  {author} {\bibinfo {author} {\bibfnamefont {A.}~\bibnamefont
  {Cavagna}},\ }\href@noop {} {\bibfield  {journal} {\bibinfo  {journal} {Phys.
  Rep.}\ }\textbf {\bibinfo {volume} {476}},\ \bibinfo {pages} {51} (\bibinfo
  {year} {2009})}\BibitemShut {NoStop}%
\bibitem [{\citenamefont {Castellani}\ and\ \citenamefont
  {Cavagna}(2005)}]{castellani05_spin-glass_pedestrians}%
  \BibitemOpen
  \bibfield  {author} {\bibinfo {author} {\bibfnamefont {T.}~\bibnamefont
  {Castellani}}\ and\ \bibinfo {author} {\bibfnamefont {A.}~\bibnamefont
  {Cavagna}},\ }\href@noop {} {\bibfield  {journal} {\bibinfo  {journal} {J.
  Stat. Mech.: Theor. Exp.}\ } (\bibinfo {year} {2005})}\BibitemShut {NoStop}%
\bibitem [{Note1()}]{Note1}%
  \BibitemOpen
  \bibinfo {note} {Dephasing dynamics was early studied in connection with the
  mean field approximation to the non-equilibrium dynamics of isolated
  systems~\cite {habib96_dephasing_mean_field}. In such case, the quasifree
  modes are the modes diagonalizing the quadratic mean field
  Hamiltonian.}\BibitemShut {Stop}%
\bibitem [{\citenamefont {F\"{u}rst}\ \emph {et~al.}(2013)\citenamefont
  {F\"{u}rst}, \citenamefont {Mendl},\ and\ \citenamefont
  {Spohn}}]{furst13_qbe_hubbard_1d_nnn}%
  \BibitemOpen
  \bibfield  {author} {\bibinfo {author} {\bibfnamefont {M.}~\bibnamefont
  {F\"{u}rst}}, \bibinfo {author} {\bibfnamefont {C.}~\bibnamefont {Mendl}}, \
  and\ \bibinfo {author} {\bibfnamefont {H.}~\bibnamefont {Spohn}},\
  }\href@noop {} {\bibfield  {journal} {\bibinfo  {journal} {Phys. Rev. E}\
  }\textbf {\bibinfo {volume} {88}},\ \bibinfo {pages} {012108} (\bibinfo
  {year} {2013})}\BibitemShut {NoStop}%
\bibitem [{Note2()}]{Note2}%
  \BibitemOpen
  \bibinfo {note} {In the particular case of $J_2=0$ we also find short time
  plateaus, but the distance saturate to a non-zero constant.}\BibitemShut
  {Stop}%
\bibitem [{\citenamefont {Erdos}\ \emph {et~al.}(2004)\citenamefont {Erdos},
  \citenamefont {Salmhofer},\ and\ \citenamefont {Yau}}]{erdos04_qbe}%
  \BibitemOpen
  \bibfield  {author} {\bibinfo {author} {\bibfnamefont {L.}~\bibnamefont
  {Erdos}}, \bibinfo {author} {\bibfnamefont {M.}~\bibnamefont {Salmhofer}}, \
  and\ \bibinfo {author} {\bibfnamefont {H.-T.}\ \bibnamefont {Yau}},\
  }\href@noop {} {\bibfield  {journal} {\bibinfo  {journal} {J. Stat. Phys.}\
  }\textbf {\bibinfo {volume} {116}},\ \bibinfo {pages} {367} (\bibinfo {year}
  {2004})}\BibitemShut {NoStop}%
\bibitem [{\citenamefont {Berges}\ and\ \citenamefont
  {Bors\'{a}nyi}(2006)}]{berges06_validity_transp_eqs}%
  \BibitemOpen
  \bibfield  {author} {\bibinfo {author} {\bibfnamefont {J.}~\bibnamefont
  {Berges}}\ and\ \bibinfo {author} {\bibfnamefont {S.}~\bibnamefont
  {Bors\'{a}nyi}},\ }\href@noop {} {\bibfield  {journal} {\bibinfo  {journal}
  {Phys. Rev. D}\ }\textbf {\bibinfo {volume} {74}},\ \bibinfo {pages} {045022}
  (\bibinfo {year} {2006})}\BibitemShut {NoStop}%
\bibitem [{\citenamefont {Habib}\ \emph {et~al.}(1996)\citenamefont {Habib},
  \citenamefont {Kluger}, \citenamefont {Mottola},\ and\ \citenamefont
  {Paz}}]{habib96_dephasing_mean_field}%
  \BibitemOpen
  \bibfield  {author} {\bibinfo {author} {\bibfnamefont {S.}~\bibnamefont
  {Habib}}, \bibinfo {author} {\bibfnamefont {Y.}~\bibnamefont {Kluger}},
  \bibinfo {author} {\bibfnamefont {E.}~\bibnamefont {Mottola}}, \ and\
  \bibinfo {author} {\bibfnamefont {J.~P.}\ \bibnamefont {Paz}},\ }\href@noop
  {} {\bibfield  {journal} {\bibinfo  {journal} {Phys. Rev. Lett.}\ }\textbf
  {\bibinfo {volume} {76}},\ \bibinfo {pages} {4660} (\bibinfo {year}
  {1996})}\BibitemShut {NoStop}%
\bibitem [{Note3()}]{Note3}%
  \BibitemOpen
  \bibinfo {note} {We came to know about this reference after developing the
  derivation independently.}\BibitemShut {Stop}%
\bibitem [{\citenamefont {Linz}(1985)}]{linz85_volterra_eqs}%
  \BibitemOpen
  \bibfield  {author} {\bibinfo {author} {\bibfnamefont {P.}~\bibnamefont
  {Linz}},\ }\href@noop {} {\emph {\bibinfo {title} {Analytical and numerical
  methods for Volterra equations}}}\ (\bibinfo  {publisher} {SIAM},\ \bibinfo
  {year} {1985})\BibitemShut {NoStop}%
\end{thebibliography}

%

%
%
%

\newpage


\setcounter{equation}{0}
\setcounter{figure}{0}
\onecolumngrid

{\Large \center {\bf Supplementary Material to: Glass-like Behavior in a System of One Dimensional Fermions after a Quantum Quench}} \\

In this supplement we will show the derivation of the evolution equations used to extract the results discussed in the main text. We will make a brief outline of the derivation of the evolution equation for the momentum distribution since it has been already been presented in full detail in Ref.~\cite{nessi14_evolution_eqs_pot}, and pay most of the attention to the evolution equation for the fluctuations. We also include a discussion on the behavior of the relaxation timescales with system size.

\section{Evolution equation for the momentum distribution}
We shall first make some elemental definitions to set up the situation. We consider a system of interacting spinless fermions with Hamiltonian $H=H_0+\alpha H_1$ with
\begin{equation}
H=H_0+\alpha H_1=\sum_{\mathbf{k}}\epsilon(\mathbf{k})n(\mathbf{k})+\alpha\sum_{\mathbf{k}_{1},\mathbf{k}_{2},\mathbf{k}_{3},\mathbf{k}_{4}}V^{\mathbf{k}_{1},\mathbf{k}_{2}}_{\mathbf{k}_{3},\mathbf{k}_{4}}c^{\dagger}(\mathbf{k}_{1})c^{\dagger}(\mathbf{k}_{2})c(\mathbf{k}_{3})c(\mathbf{k}_{4}),
\end{equation}
where $c^{\dagger}(\mathbf{k})$ and $c(\mathbf{k})$ are fermionic creation and annihilation operators satisfying canonical anticommutation relations, $V^{\mathbf{k}_{1},\mathbf{k}_{2}}_{\mathbf{k}_{3},\mathbf{k}_{4}}$ is the momentum-space matrix element of the interaction, $\epsilon(\mathbf{k})$ is the dispersion relation, $n(\mathbf{k})=c^{\dagger}(\mathbf{k})c(\mathbf{k})$ is the number operator and $\alpha$ is the strength of the interaction. Our results can be easily extended to the bosonic case. The hermiticity of the Hamiltonian and the symmetry in the sum indices impose $V^{\mathbf{k}_{1},\mathbf{k}_{2}}_{\mathbf{k}_{3},\mathbf{k}_{4}}=-V^{\mathbf{k}_{2},\mathbf{k}_{1}}_{\mathbf{k}_{3},\mathbf{k}_{4}}=-V^{\mathbf{k}_{1},\mathbf{k}_{2}}_{\mathbf{k}_{4},\mathbf{k}_{3}}=\bar{V}^{\mathbf{k}_{4},\mathbf{k}_{3}}_{\mathbf{k}_{2},\mathbf{k}_{1}}$, where $\bar{V}$ denotes the complex conjugate.

Furthermore, we will be interested in the special case of a translationally invariant Hamiltonian in which the particles interact via a pair potential $v(\mathbf{x}-\mathbf{y})$. In such case
\begin{equation}
V^{\mathbf{k}_{1},\mathbf{k}_{2}}_{\mathbf{k}_{3},\mathbf{k}_{4}}=\frac{1}{4V}\delta_{\mathbf{k}_{1}+\mathbf{k}_{2},\mathbf{k}_{3}+\mathbf{k}_{4}}\left(\hat{v}(\mathbf{k}_{1}-\mathbf{k}_{4})-\hat{v}(\mathbf{k}_{2}-\mathbf{k}_{4})-\hat{v}(\mathbf{k}_{1}-\mathbf{k}_{3})+\hat{v}(\mathbf{k}_{2}-\mathbf{k}_{3})\right),
\end{equation}
 where $\hat{v}(\mathbf{k})$ is the Fourier transform of the potential and we have written the antisymmetrized version in order to respect the symmetry conditions of the potential. We are interested in the evolution of the system starting from an arbitrary initial condition given by a density matrix $\rho(0)$ which we leave unspecified for the moment.

In the specific $1D$ model treated in the main text $\epsilon(k)=-2 J_1 \cos(2k\pi/L)-2 J_2 \cos(4 k\pi/L)$ and $\hat{v}(k-k')=[\Delta_{1}e^{i(k'-k)2\pi/L}+\Delta_{2}e^{i(k'-k)4\pi/L}]$.

To obtain an evolution equation for the momentum distribution we start from the Liouville equation in the interaction representation ($\hbar$=1):
\begin{equation}\label{eq:liouville}
\partial_{t}\tilde{\rho}(t)=-i\alpha[\tilde{H}_{1}(t),\tilde{\rho}(t)]=\alpha L(t)\tilde{\rho}(t),
\end{equation}
where $\tilde{O}(t)=e^{iH_{0}t}Oe^{-iH_{0}t}$ is the interaction representation of the operator $O$ and we have introduced the Liouville superoperator $L(t)O=-i[\tilde{H}_{1}(t),O]$. Our task is to find approximate solutions to the microscopic dynamics described by the Liouville equation. The POT defines a program for achieving this. We need to first identify the ``slow'' or ``macroscopic'' variables in our system and then project the dynamics into the subspace of these slow variables. As noticed in the main text, in a weakly interacting homogeneous system the occupation number operators emerge as natural slow variables since $[H,n(\mathbf{k})]=\mathcal{O}(\alpha)$. To perform the projection we first introduce the ``relevant'' density matrix
\begin{equation}
\sigma(t)=\frac{1}{Z(t)}\exp\left[-\sum_{\mathbf{k}}\lambda(\mathbf{k},t)n(\mathbf{k})\right],
\end{equation}
where the time-dependent partition function is given by $Z(t)=\mathrm{Tr}\left[\exp\left(-\sum_{\mathbf{k}}\lambda(\mathbf{k},t)n(\mathbf{k})\right)\right]$. Note that $\tilde{\sigma}(t)=\sigma(t)$. The Lagrange multipliers $\lambda(\mathbf{k},t)$ enforce the relation:
\begin{equation}\label{eq:constraints}
\langle n(\mathbf{k})\rangle_{t}\equiv \mathrm{Tr}[n(\mathbf{k})\sigma(t)]=\mathrm{Tr}[n(\mathbf{k})\rho(t)].
\end{equation}

The projection of the dynamics consists in finding an equation of motion for $\sigma(t)$. To this end we introduce a projection super-operator $P(t)$ that projects the relevant density matrix $P(t)\tilde{\rho}(t)=\tilde{\sigma}(t)$:
\begin{equation}\label{eq:projector}
P(t)\mu=\left(\sigma(t)-\sum_{\mathbf{k}}\frac{\delta\sigma(t)}{\delta\langle n(\mathbf{k})\rangle_{t}}\langle n(\mathbf{k})\rangle_{t}\right)\mathrm{Tr}\left[\mu\right]+\sum_{\mathbf{k}}\frac{\delta\sigma(t)}{\delta\langle n(\mathbf{k})\rangle_{t}}\mathrm{Tr}\left[n(\mathbf{k})\mu\right],
\end{equation}
where $\mu$ is an arbitrary density matrix. The projection operator~\eqref{eq:projector} is specially designed to satisfy the following properties~\cite{grabert82_book_pot,rau95_review_pot,breuer82_book_open_systems}:
\begin{eqnarray}\label{eq:p_properties}
\nonumber P(t)\tilde{\rho}(t)&=&\tilde{\sigma}(t),\\
\nonumber P(t)\partial_{t}\tilde{\rho}(t)&=&\partial_{t}\tilde{\sigma}(t),\\
\nonumber \mathrm{Tr}\left[n(\mathbf{k})P(t)\mu\right]&=&\mathrm{Tr}\left[n(\mathbf{k})\mu\right],
\nonumber \\P(t)P(t')\mu&=&P(t)\mu,\\
P(t)L(t)P(s)\mu&=&0.
\end{eqnarray}
The fourth identity, setting $t=t'$, expresses the idempotent character of the projector, while the last identity depends on the explicit form of the Hamiltonian $H$, in particular, on momentum conservation.
It is also useful to define the complementary projector $Q(t)=1-P(t)$.

Following the usual steps~\cite{grabert82_book_pot,rau95_review_pot,breuer82_book_open_systems,robertson66_pot_first}, introducing projectors in the Liouville equation, we obtain an equation for the dynamics of the slow degrees of freedom
\begin{equation}
\partial_{t}P(t)\tilde{\rho}(t)=\alpha P(t)L(t)\tilde{\rho}(t),
 \end{equation}
and other for the fast, microscopic degrees of freedom
\begin{equation}\label{eq:fast_dof}
\partial_{t}Q(t)\tilde{\rho}(t)=\alpha Q(t)L(t)\tilde{\rho}(t).
\end{equation}
Inserting the identity $I=P(t)+Q(t)$ in both equations we obtain the system:
\begin{eqnarray}\label{eq:fast_slow}
\partial_{t}P(t)\tilde{\rho}(t)&=&\alpha P(t)L(t)P(t)\tilde{\rho}(t)+\alpha P(t)L(t)Q(t)\tilde{\rho}(t),\\
\partial_{t}Q(t)\tilde{\rho}(t)&=&\alpha Q(t)L(t)P(t)\tilde{\rho}(t)+\alpha Q(t)L(t)Q(t)\tilde{\rho}(t).
\end{eqnarray}
The equation for the relevant density matrix $\sigma(t)$ can be obtained solving the equation for the irrelevant part $Q(t)\tilde{\rho}(t)$ in the second line of the system and inserting the solution in the first line. The second line is a linear first order homogeneous differential equation in the operator $Q(t)\tilde{\rho}(t)$ (the inhomogeneity is $\alpha Q(t)L(t)P(t)\tilde{\rho}(t)$) that can be (formally) solved in the same way as a real valued function differential equation. The solution is:
\begin{equation}\label{eq:formal_sol}
Q(t)\tilde{\rho}(t)=\alpha\int_{0}^{t}ds\, G(t,s)Q(s)L(s)P(s)\tilde{\rho}(s)+G(0,t)Q(0)\tilde{\rho}(0),
\end{equation}
where $G(t,s)$ is an ordered exponential $G(s,t)=\mathrm{T_{\rightarrow}}\exp\left[-\alpha\int_{s}^{t}ds'\, Q(s')L(s')\right]$, i.e., the solution of the equation
\begin{eqnarray}
\partial_{t}G(s,t)&=&-\alpha G(s,t)Q(t)L(t),\\
G(s,s)&=&I.
\end{eqnarray}

Inserting~\eqref{eq:formal_sol} in the first line of~\eqref{eq:fast_slow} we obtain the desired equation:
 \begin{equation}\label{eq:robertson}
 \partial_{t}\tilde{\sigma}(t)=\alpha P(t)L(t)\tilde{\sigma}(t)+\alpha^{2}\int_{0}^{t}ds\, P(t)L(t)G(t,s)Q(s)L(s)\tilde{\sigma}(s)+\alpha P(t)L(t)G(t,0)Q(0)\tilde{\rho}(0),
 \end{equation}
The first term in Eq.~\eqref{eq:robertson} is a mean field-like term that vanish due to momentum conservation, the second one can be expressed entirely in terms of the past history of the momentum distribution $\langle n(\mathbf{k})\rangle_{t}$, and the third one is a microscopic noise that can not be expressed in terms of the slow variables. The last term in Eq.~\eqref{eq:robertson} (the microscopic noise) disappears if we chose an initial condition of the same form of the relevant density matrix, i.e., if $\rho(0)=\sigma(0)$. We shall then chose Gaussian (uncorrelated) initial density matrices, such as the ground state of $H_0$ or a finite temperature state. We are thus considering an interaction quench.

 Eq.~\eqref{eq:robertson} is equivalent to the Liouville dynamics and, in general, as difficult to solve as the original problem. It sets, however, a good starting point for approximations. To render Eq.~\eqref{eq:robertson} tractable we perform a perturbative expansion in the interaction strength using that $G(t,s)=I+\mathcal{O}(\alpha)$. Taking the trace $\langle n(\mathbf{k})\rangle_{t}=\mathrm{Tr}[n(\mathbf{k})\sigma(t)]$ we finally obtain
 \begin{equation}\label{eq:pert_rob}
 \partial_{t}\langle n(\mathbf{k})\rangle_{t}=\alpha^{2}\int_{0}^{t}ds\,\mathrm{Tr}\left[n(\mathbf{k})L(t)L(s)\tilde{\sigma}(s)\right]+\mathcal{O}(\alpha^{3}).
 \end{equation}
A great simplification arises since, given the Gaussian structure of $\sigma(t)$, we can use the Wick pairing rule to evaluate the trace in~\eqref{eq:pert_rob}. After a straightforward (but potentially tedious) calculation we obtain the explicit equation of motion
\begin{eqnarray}\label{eq:kinetic}
\nonumber f(\mathbf{k},t)&=&f(\mathbf{k},0)-16\alpha^{2}\sum_{\mathbf{k}_{2},\mathbf{k}_{3},\mathbf{k}_{4}}\vert V^{\mathbf{k},\mathbf{k}_{2}}_{\mathbf{k}_{3},\mathbf{k}_{4}}\vert^{2}\int_{0}^{t}ds\,\frac{\sin\left[(t-s)\Delta e^{\mathbf{k},\mathbf{k}_2}_{\mathbf{k}_3,\mathbf{k}_4}\right]}{\Delta e^{\mathbf{k},\mathbf{k}_2}_{\mathbf{k}_3,\mathbf{k}_4}}\\
&\times&\left(f(\mathbf{k},s)f(\mathbf{k}_{2},s)\bar{f}(\mathbf{k}_{3},s)\bar{f}(\mathbf{k}_{4},s)-f(\mathbf{k}_{3},s)f(\mathbf{k}_{4},s)\bar{f}(\mathbf{k},s)\bar{f}(\mathbf{k}_{2},s)\right)+\mathcal{O}(\alpha^{3}),
\end{eqnarray}
where $\Delta e^{\mathbf{k},\mathbf{k}_2}_{\mathbf{k}_3,\mathbf{k}_4}=\epsilon(\mathbf{k})+\epsilon(\mathbf{k}_{2})-\epsilon(\mathbf{k}_{3})-\epsilon(\mathbf{k}_{4})$ and, in order to ease the notation, we have defined $f(\mathbf{k},t)\equiv\langle n(\mathbf{k})\rangle_{t}$ and $\bar{f}(\mathbf{k},t)\equiv 1-\langle n(\mathbf{k})\rangle_{t}$. This equation, in slightly different versions, has appeared many times in the literature. In Ref.~\cite{rau95_review_pot} it was derived using the same tools that we present here~\footnote{We came to know about this reference after developing the derivation independently.} but it was used only as an intermediate step to derive the Boltzmann equation whereas in Refs.~\cite{stark13_kinetic_description,erdos04_qbe} it was heuristically derived and used to study the dynamics of infinite dimensional models and to derive a quantum version of the Boltzmann equation, respectively. Eq.~\eqref{eq:kinetic} is valid for systems in the continuum limit and also for lattice systems which only conserve quasi-momentum. A discussion on the accuracy of Eq.~\eqref{eq:kinetic} can be found in Ref.~\cite{nessi14_evolution_eqs_pot}.

It is worth noticing that Eq.~\eqref{eq:kinetic} includes kinetic collisions ($\Delta e^{\mathbf{k},\mathbf{k}_2}_{\mathbf{k}_3,\mathbf{k}_4}=0$) as well as non-kinetic processes. This is related with the fact that Eq.~\eqref{eq:kinetic} describes the dynamics of the system in all timescales. For short timescales, where there has not yet elapsed enough time for particles to collide, non-kinetic processes dominate the dynamics, whereas for long timescales kinetic collisions are dominant.

With respect to implementation details, Eq.~\eqref{eq:kinetic} can be solved using standard techniques for systems of Volterra integral equations~\cite{linz85_volterra_eqs}. A straightforward algorithm for the solution using, for instance, the trapezoidal rule to perform the time integral, implies a calculation time that scales as $L^{3D}\times N^2$, where $N$ is the number of times steps and $D$ the space dimension. We have found an algorithm whose execution time scales as $L^{3D}\times N$ allowing us to reach large sizes and times. We finally note that the evolution equations are very suitable for parallel computing.

\section{Evolution equation for higher order correlations}

In this section we undertake the calculation of two-times correlation functions of the slow variables, often referred as the fluctuations of the variable. We will briefly review the projection operator formalism of~\cite{grabert82_book_pot}. To perform the calculation we need to switch to the Heisenberg representation, where an operator with no explicit tome-dependence evolves according to the law:
\begin{equation}
  \partial_{t}O=i[H,O]\equiv iLO.
\end{equation}
Note the difference with the Liouvillian defined in~\eqref{eq:liouville}. The trace operation defines a dual projection operator over the observables of the Hilbert space:
\begin{equation}
\mathrm{Tr}\left[OP(t)\mu\right]=\mathrm{Tr}\left[\mu\mathsf{P}(t)O\right],
\end{equation}
where $O$  is an observable, $\mu$  a density matrix and $\mathsf{P}(t)$  is the observable space projection operator. Its explicit form can be obtained right from the last expression and reads in our case,
\begin{equation}\label{eq:heisenberg_proj}
\mathsf{P}(t)O=\mathrm{Tr}[\sigma(t)O]+\sum_{\vec{k}}(n(\vec{k})-\langle n(\vec{k})\rangle_{t})\mathrm{Tr}\left[\frac{\delta\sigma(t)}{\delta\langle n(\vec{k})\rangle_{t}}O\right].
\end{equation}
It can be readily shown that this dual projector satisfies the convenient properties:
\begin{eqnarray}\label{eq:prop_heisenber}
\mathsf{P}(t)\mathsf{P}(t')&=&\mathsf{P}(t'),\\
\dot{\mathsf{P}}(t)&=&\mathsf{P}(t)\dot{\mathsf{P}}(t)(1-\mathsf{P}(t)).
\end{eqnarray}
It is also useful to define the complementary projector $\mathsf{Q}(t)=1-\mathsf{P}(t)$.

In the Heisenberg representation the strategy is to separate the slow and fast components of the evolution operator $e^{iLt}$ using the projector $\mathsf{P}(t)$ and its complement $\mathsf{Q}(t)$:
\begin{equation}
e^{iLt}=e^{iLt}\mathsf{P}(t)+e^{iLt}\mathsf{Q}(t).
\end{equation}
Using the last identity in Eq.~\eqref{eq:prop_heisenber} we obtain the equation for the irrelevant part
\begin{equation}
\partial_{t}e^{iLt}\mathsf{Q}(t)-e^{iLt}\mathsf{Q}(t)iL\mathsf{Q}(t)=e^{iLt}\mathsf{P}(t)\left[iL-\dot{\mathsf{P}}(t)\right]\mathsf{Q}(t),
\end{equation}
that can be solved formally using the propagator $\bar{\mathsf{G}}(s,t)$  satisfying
\begin{eqnarray}
\partial_{t}\bar{\mathsf{G}}(s,t)&=&-iL\mathsf{Q}(t)\bar{\mathsf{G}}(s,t),
\\ \bar{\mathsf{G}}(s,s)&=&I,
\end{eqnarray}
i.e., the chronologically ordered exponential
\begin{equation}
\bar{\mathsf{G}}(s,t)=T_{\leftarrow}\exp\left[-\int_{s}^{t}ds'\, iL\mathsf{Q}(s')\right].
\end{equation}
The solution is
\begin{equation}
e^{iLt}\mathsf{Q}(t)=e^{iLs}\mathsf{Q}(s)\bar{\mathsf{G}}(t,s)+\int^{t}_{s}du\,e^{iLu}\mathsf{P}(u)\left[iL-\dot{\mathsf{P}}(u)\right]\mathsf{Q}(u)\bar{\mathsf{G}}(t,u).
\end{equation}
Inserting the formal solution into the relevant part we obtain
\begin{eqnarray}
e^{iLt}&=&e^{iLt}\mathsf{\mathsf{P}}(t)+\int_{s}^{t}du\, e^{iLu}\mathsf{\mathsf{P}}(u)(iL-\dot{\mathsf{\mathsf{P}}}(u))(1-\mathsf{\mathsf{P}}(u))\bar{\mathsf{G}}(t,u)
\\&&+e^{iLs}(\mathsf{1-\mathsf{P}}(s))\bar{\mathsf{G}}(t,s),
\end{eqnarray}
where $s$  is an arbitrary time $0\leq s\leq t$. Defining a new ordered exponential

\begin{equation}
\mathsf{G}(s,t)=T_{\leftarrow}\exp\left[\int_{s}^{t}ds'\, iL\mathsf{Q}(s')\right],
\end{equation}
we can write
\begin{eqnarray}\label{eq:ev_split}
e^{iLt}&=&e^{iLt}\mathsf{\mathsf{P}}(t)+\int_{s}^{t}du\, e^{iLu}\mathsf{P}(u)(iL-\dot{\mathsf{P}}(u))(1-\mathsf{P}(u))\mathsf{G}(u,t)\\
&&+e^{iLs}(\mathsf{1-\mathsf{P}}(s))\mathsf{G}(s,t).
\end{eqnarray}
Using the explicit form of the projector superoperator Eq.~\ref{eq:heisenberg_proj}, it is possible to obtain an operator Langevin-like equation for the slow variables~\cite{grabert82_book_pot},
\begin{eqnarray}\label{eq:langevine}
\dot{n}(\vec{k},t)&\equiv e^{iLt}\dot{n}(\vec{k})&=v_{\vec{k}}(t)+\sum_{\vec{k}'}\Omega_{\vec{k},\vec{k}'}(t)\,\delta n(\vec{k},t)+\\
&&+\int_{s}^{t}du\left(K_{\vec{k}}(t,u)+\sum_{\vec{k}'}\Phi_{\vec{k},\vec{k}'}(t,u)\:\delta n(\vec{k}',u)\right)\\
&&+\eta_{\vec{k}}(t,s).
\end{eqnarray}
We have defined
\begin{eqnarray}
\dot{n}(\vec{k})&=&iLn(\vec{k}),\\
\delta n(\vec{k},t)&=&n(\vec{k},t)-\langle n(\vec{k})\rangle_{t},
\end{eqnarray}
where $n(\vec{k},t)=e^{iLt}n(\vec{k})$ . The organized drift $v_{\vec{k}}(t)=\mathrm{Tr}\left\{ n(\vec{k})[H,\sigma(t)]\right\}$   and the collective frequencies $\Omega_{\vec{k},\vec{k}'}(t)=\mathrm{Tr}\left[\frac{\delta\sigma(t)}{\delta\langle n(\vec{k})\rangle_{t}}\dot{n}(\vec{k})\right]=\frac{\delta v_{\vec{k}}(t)}{\delta\langle n(\vec{k})\rangle_{t}}$ , vanish identically due to momentum conservation in $H_1$. The after effect functions $K_{\vec{k}}(t,u)$ can be expressed in terms of the $\langle n(\mathbf{k})\rangle_{t}$'s:
\begin{equation}\label{eq:after_effect}
K_{\vec{k}}(t,u)=\mathrm{Tr}\left[\sigma(u)iL\mathsf{Q}(u)\mathsf{G}(u,t)\dot{n}(\vec{k})\right]=-\alpha^{2}\mathrm{Tr}\left\{ [H_{1}(u),[H_{1}(t),n(\vec{k})]]\sigma(u)\right\} +\mathcal{O}(\alpha^{3}),
\end{equation}
and are related with the dynamics of the momentum distribution, see Eq.~\eqref{eq:pert_rob}. In the last equality of Eq.~\eqref{eq:after_effect} we have used  $\mathsf{G}(u,t)=I+\mathcal{O}(\alpha)$. The memory functions $\Phi_{\vec{k},\vec{k}'}(u,t)$ read
\begin{eqnarray}
\Phi_{\vec{k},\vec{k}'}(u,t)&=&\mathrm{Tr}\left[\frac{\delta\sigma(u)}{\delta\langle n(\vec{k}')\rangle_{u}}iL\mathsf{Q}(u)\mathsf{G}(u,t)\dot{n}(\vec{k})\right]\\
&&-\sum_{\vec{k}''}\langle n(\vec{k}'')\rangle_{u}\mathrm{Tr}\left[\frac{\delta^{2}\sigma(u)}{\delta\langle n(\vec{k}')\rangle_{u}\delta\langle n(\vec{k}'')\rangle_{u}}\mathsf{G}(u,t)\dot{n}(\vec{k})\right].
\end{eqnarray}
This functions are related to the after effect functions via a functional derivative~\cite{grabert82_book_pot}
\begin{equation}
\Phi_{\vec{k},\vec{k}'}(t,u)=\frac{\delta\int_{0}^{t}ds\, K_{\vec{k}}(t,s)}{\delta\langle n(\vec{k}')\rangle_{u}},
\end{equation}
which constitutes the key relation between the dynamics of the momentum distribution and the fluctuations. Lastly, we have defined the microscopic noise
\begin{equation}
\eta_{\vec{k}}(t,s)=e^{iLs}(\mathsf{1-\mathsf{P}}(s))\mathsf{G}(s,t)\dot{n}(\vec{k}).
\end{equation}

If we take the trace with respect to $\rho(0)$ in the Eq.~\eqref{eq:langevine} and we keep only with the lowest order in $\alpha$ we will recover the kinetic equation for the momentum distribution Eq.~\eqref{eq:kinetic}. But our intention is to rederive that equation but to make approximations on the dynamics of the operators themselves in order to calculate higher order correlation functions.

Setting $s=0$  in the Langevin Eq.~\eqref{eq:langevine} and subtracting the mean value we obtain an equation for the fluctuations:
\begin{equation}\label{eq:true_langevin}
\partial_{t}\delta{n}(\mathbf{k},t)=\int_{0}^{t}ds\,\sum_{\mathbf{k}'}\Phi_{\mathbf{k},\mathbf{k}'}(s,t)\:\delta n(\mathbf{k}',s)+\eta_{\mathbf{k}}(t),
\end{equation}
where $\delta n(\mathbf{k},t)=e^{iLt}n(\mathbf{k})-\langle n(\mathbf{k})\rangle_{t}$ and

\begin{equation}
\eta_{\vec{k}}(t)=\eta_{\vec{k}}(t,0)-\mathrm{Tr}\left[\rho(0)\eta_{\vec{k}}(t,0)\right]=\eta_{\vec{k}}(t,0).
\end{equation}
From this definition of the noise is clear that $\langle\eta_{\vec{k}}(t)\rangle=\mathrm{Tr}\left[\rho(0)\eta_{\vec{k}}(t)\right]=0$.

Starting from Eq.~\eqref{eq:true_langevin}, using the Wick rule and taking the functional derivative in Eq.~\eqref{eq:after_effect} we find an explicit evolution equation for the time-correlation function of the slow variables $F_{\mathbf{k},\mathbf{k}'}(t)\equiv\mathrm{Tr}\left[\rho(0)\delta n(\mathbf{k},t)\delta n(\mathbf{k}',0)\right]$:
\begin{equation}\label{eq:fluc}
F_{\mathbf{k},\mathbf{k}'}(t)=F_{\mathbf{k},\mathbf{k}'}(0)-16\alpha^{2}\int_{0}^{t}ds\,\left\{ F_{\mathbf{k},\mathbf{k}'}(s)A_{\mathbf{k}}(t,s)+\sum_{\mathbf{q}}F_{\mathbf{q},\mathbf{k}'}(s)\left[B_{\mathbf{k},\mathbf{q}}(t,s)-2C_{\mathbf{k},\mathbf{q}}(t,s)\right]\right\},
\end{equation}
where the matrices can be written as (using the notation defined earlier)
\begin{eqnarray}
\nonumber B_{\mathbf{k},\mathbf{q}}(t,s)&=&\sum_{\mathbf{k}_{3},\mathbf{k}_{4}}\vert V^{\mathbf{k},\mathbf{q}}_{\mathbf{k}_{3},\mathbf{k}_{4}}\vert^{2}\frac{\sin\left[(t-s)\Delta e^{\mathbf{k},\mathbf{q}}_{\mathbf{k}_3,\mathbf{k}_4}\right]}{\Delta e^{\mathbf{k},\mathbf{q}}_{\mathbf{k}_3,\mathbf{k}_4}}\left(f(\mathbf{k},s)\bar{f}(\mathbf{k}_{3},s)\bar{f}(\mathbf{k}_{4},s)+\bar{f}(\mathbf{k},s)f(\mathbf{k}_{3},s)f(\mathbf{k}_{4},s)\right),\\
\nonumber C_{\mathbf{k},\mathbf{q}}(t,s)&=&\sum_{\mathbf{k}_{2},\mathbf{k}_{4}}\vert V^{\mathbf{k},\mathbf{k}_{2}}_{\mathbf{q},\mathbf{k}_{4}}\vert^{2}\frac{\sin\left[(t-s)\Delta e^{\mathbf{k},\mathbf{k}_2}_{\mathbf{q},\mathbf{k}_4}\right]}{\Delta e^{\mathbf{k},\mathbf{k}_2}_{\mathbf{q},\mathbf{k}_4}}\left(f(\mathbf{k},s)f(\mathbf{k}_{2},s)\bar{f}(\mathbf{k}_{4},s)+\bar{f}(\mathbf{k},s)\bar{f}(\mathbf{k}_{2},s)f(\mathbf{k}_{4},s)\right),
\end{eqnarray}
and $A_{\mathbf{k}}(t,s)=\sum_{\mathbf{k}'}B_{\mathbf{k}',\mathbf{k}}(t,s)$. To arrive to Eq.~\eqref{eq:fluc} we have to take into account that $\Phi_{\vec{k},\vec{k}'}(t,s)=\mathcal{O}(\alpha^2)$ and that $\mathsf{G}(s,t)=I+\mathcal{O}(\alpha)$. Notice that the matrices satisfy the convenient property $A_{\vec{k}}(t,t)=B_{\vec{k},\vec{q}}(t,t)=C_{\vec{k},\vec{q}}(t,t)=0$ reflecting causality, i.e., the value of the fluctuations at time $t$ only depends on the history of the fluctuations for times strictly before $t$. A similar statement can be done for the equation for the momentum distribution Eq.~\eqref{eq:kinetic}, the kernel of the integral equation vanishes for $s=t$. This property of the evolution equations brings a big technical simplification since it is not necessary to solve autoconsistent equations at each time step in the numerical integration.

To the best of our knowledge, the equation~\eqref{eq:fluc} was first presented in Ref.~\cite{nessi14_evolution_eqs_pot} and used to investigate the dynamics of a concrete system in the present publication. In order to solve Eq.~\eqref{eq:fluc} we need first to know the dynamics of the momentum distribution, i.e., calculate the solution to Eq.~\eqref{eq:kinetic}, in order to determine the coefficients $A_{\mathbf{k}}(t,s)$, $B_{\mathbf{k},\mathbf{q}}(t,s)$ and $C_{\mathbf{k},\mathbf{q}}(t,s)$. With this input, Eq.~\eqref{eq:fluc} is as amenable to numerical solution as Eq.~\eqref{eq:kinetic}. We finally remark that since the projection operator in the Heisenberg representation works directly on the evolution operator itself it would be possible to obtain similar evolution equations for other observables.

\section{Relaxation timescales and system size}

\begin{figure}
  \includegraphics[width=\largefigwidth]{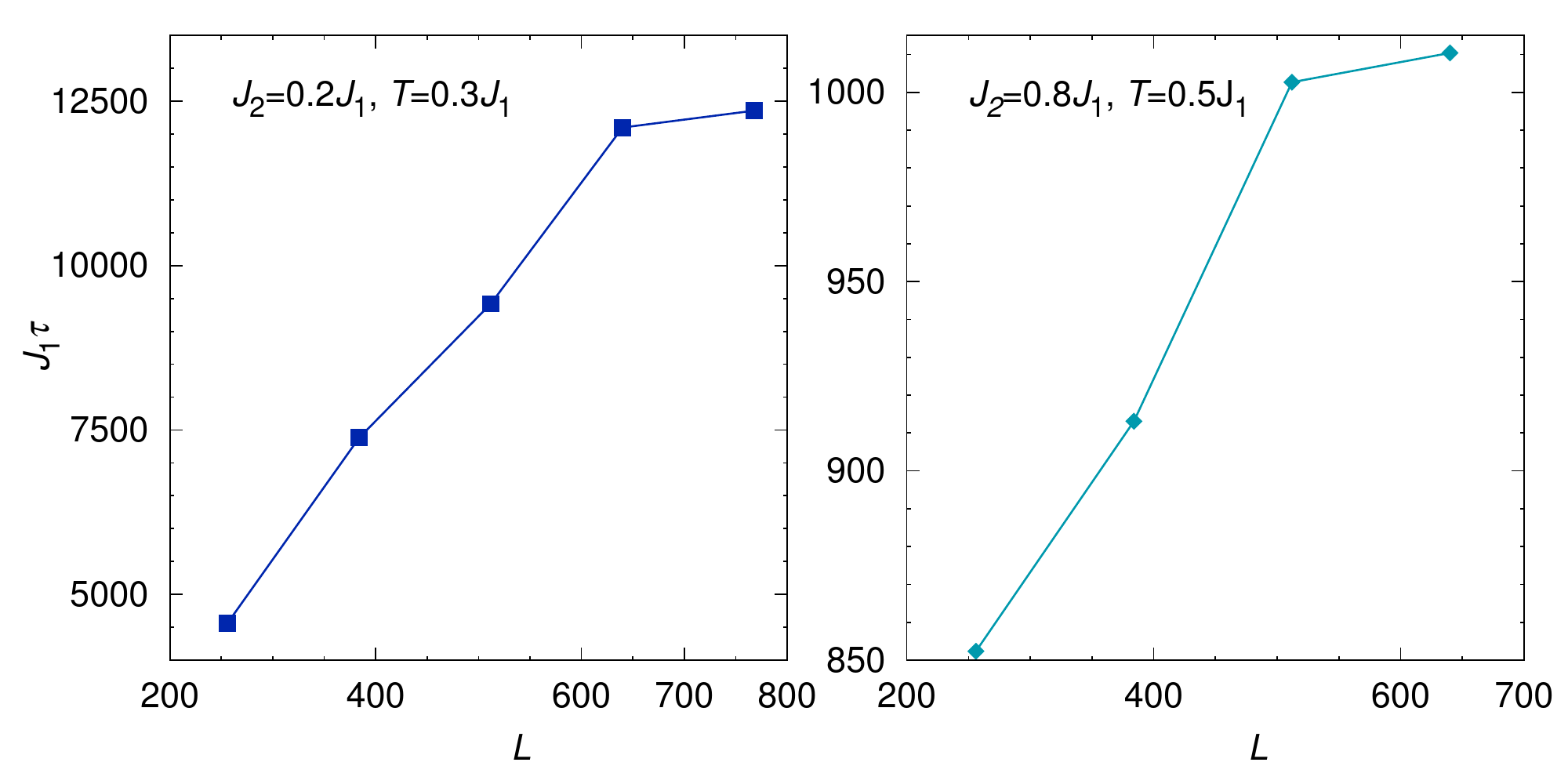}\\
  \caption{Relaxation timescale of the correlator $C_J(t)$ as a function of the size of the system. Left panel: For $J_2=0.2J_1$ and $T=0.3J_1$, deep in the parameter region with long lived prethermalized states. Right panel: For $J_2=0.8J_1$ and $T=0.5J_1$, deep in the parameter region exhibiting one-step relaxation.}\label{fig:finitesize}
\end{figure}

Having considered finite size results it is important to know about the dependence of the results on the system size, $L$. In Fig.~\ref{fig:finitesize} we show the dependence of the relaxation timescale of the correlation function $C_J(t)$ with system size for a system deep in the ``glassy'' phase and for a system with normal relaxation. We observe that the timescales of the normal system increase almost linearly with a slope $\sim 2$ until, around $L\sim500$ it begins to saturate to the thermodynamic limit value. For the glassy system the increase is again almost linear with system size with a (considerably larger) slope $\sim 20$ until, around $L\sim 700$ it begins to saturate to the thermodynamic limit value. This is another sign pointing to the fact that the relaxation mechanism in the glassy phase is qualitatively different from that of the normal phase.

\end{document}